\documentclass[12pt,preprint]{aastex}

\begin{document}

\title{ON THE PREDICTED AND OBSERVED COLOR BOUNDARIES OF THE RR LYRAE 
INSTABILITY STRIP AS A FUNCTION OF METALLICITY}
\author{ALLAN SANDAGE}
\affil{ The Observatories of the Carnegie Institution of Washington, 813 
       Santa Barbara Street, Pasadena California, 91101}

\begin{abstract}
The purpose of the paper is to predict the temperature at 
the fundamental blue edge (FBE) of the instability strip for RR 
Lyrae (RRL) variables from the pulsation equation that relates 
temperature to period, luminosity, and mass. Modern data for the 
correlations between period, luminosity, and metallicity at the 
FBE for field and cluster RRL are used for the temperature 
calculation. The predicted temperatures are changed to B-V colors 
using an adopted color transformation. The predicted temperatures 
at the FBE become hotter as [Fe/H] changes from 0 to -1.5, and 
thereafter cooler as the metallicity decreases to -2.5 and 
beyond. The temperature range over this interval of metallicity 
is $\Delta$log $T_e$ = 0.04, or 640 K at 6900K. The predicted color 
variation is at the level of 0.03 mag in B-V. The predictions are 
compared with the observed RRL colors at the FBE for both the 
field and cluster variables, showing general agreement at the 
level of 0.02 mag in (B-V)$_o$, which, however, is the uncertainty 
of the reddening corrections.  

The focus of the problem is then reversed by fitting a 
better envelope to the observed FBE relation between color and 
metallicity for metallicities smaller than -1.8 which, when 
inserted in the pulsation equation, gives a non-linear 
calibration of the absolute magnitude of the average evolved 
level of the HB of M$_V$ = 1.109 + 0.600 ([Fe/H]) + 0.140 
([Fe/H])$^2$, where the zero point has been set by the observed 
RR Lyrae stars in the LMC at $\langle$V$\rangle_o$ = 19.064 for [Fe/H] = -1.5, 
and using an adopted LMC modulus of (m-M)$_o$ = 18.54.      

The position of the envelope locus at the shortest periods 
for the observed period-metallicity correlation differs between 
the field and cluster variables by $\Delta$log P = 0.029 $\pm$ 0.007,
the field variables having the shorter periods at the envelope. The 
field and cluster variables also differ in the near absence of 
cluster RR Lyraes in the -1.7 $>$ [Fe/H] $>$ -2.0 metallicity 
interval, whereas the field variables show no such gap. We aver 
that these differences require different origins for the field 
and the cluster variables. 

A test is proposed by comparing the morphology of the 
horizontal branches in the local dwarf spheroidals with that in 
the Galactic globular clusters in the inner halo and by relating 
the differences with the relevant second parameter indicators.        

\end{abstract}


\section{Introduction}

     A model was proposed at the Vatican Conference on Stellar 
Populations (O'Connell 1958) to explain the period-ratio 
dichotomy in globular cluster RR Lyrae variables, studied by 
Oosterhoff (1939, 1944), following its discovery by Grosse (1932) 
and Hachenberg (1939). The postulate of the model was that the 
difference in the observed mean periods between the two 
Oosterhoff groups could be understood if the absolute magnitude 
levels of the groups differed by $\sim$0.2 mag at constant 
temperature, the variables in the long period group being 
brighter (Sandage 1958). In the discussion, Martin Schwarzschild 
(1958) asked if, instead, the period ratio could be explained by 
a temperature difference with the long period group (Oost II) 
being cooler on average than Oosterhoff I clusters, keeping the 
luminosities the same. 

I replied that a color shift would have to be at the level 
of 0.09 mag in B-V toward the red in the long period Oosterhoff 
group. Although a later calculation (Sandage 1993b, hereafter 
S93b, footnote 1) gave 0.06 mag for the required color shift, 
both 0.06 and 0.09 mag were outside the bounds permitted by the 
observations of clusters in the two Oosterhoff period groups 
known at the time for M3 and M92 (Arp, Baum, and Sandage 1953, 
Sandage 1953) and for five other clusters (Arp 1955) over the 
relevant metallicity range of -1.5~$>$~[Fe/H]~$>$~-2.2. A near zero 
color difference of the blue fundamental edge of the strip for 
such clusters at the level of $\sim$0.02 mag in B-V was later 
confirmed from more precise photoelectric data of horizontal 
branch stars, and better determinations of the E(B-V) (Sandage 
1969).  

It was this conclusion that led to the near canonical 
assumption in the late 1980s that one should use a constant mean 
temperature of log T$_e\sim$3.85 for the mid point of the strip in 
comparing the predictions from theoretical models of the zero age 
horizontal branch (ZAHB) with the observations (S82a, S90b). 
However, this assumption of constant temperature for all 
metallicities led to the vigorous debate in the decades of 1985 
and 1995 concerning the adequacy of the extant ZAHB models or, in 
fact, whether the period shifts claimed by the observers using 
data on amplitudes were real (Caputo 1988; Lee 1990; Lee, 
Demarque, \& Zinn 1990, hereafter LDZ; Brocato, Castellani, \& 
Ripepi 1994, 1996). 

Nevertheless, {\it continuous} period shifts with metallicity, not 
just a {\it dichotomy} of mean period ratios, had been discovered by 
Preston (1959, his Fig. 4) for field RR Lyraes with the non-zero 
slope of dlogP/d([Fe/H]) $\sim$ -0.10 in the log P-metallicity 
relation. The continuum with this slope was later shown beyond 
doubt for the bright field RR Lyraes from the larger sample of 
Layden (1995, his Fig. 1). 

Other determinations of the slope at dlogP/d([Fe/H]) = 
-0.12~$\pm$~0.02 for the period-metallicity correlation had also been 
derived from a variety of independent data for RRL in clusters 
(Sandage, Katem, \& Sandage 1981, hereafter SKS; S81b, Table 7, 
S82a,b). Some of these determinations were from amplitude-
metallicity correlations, some from rise-time-metallicity 
correlations, and some from period-temperature correlations 
within the strip as a function of metallicity. 

Nevertheless, debate on the validity of period shifts 
determined using amplitudes remain. Now the focus has shifted 
from questioning the validity of the ZAHB {\it theoretical models}, to 
the role of evolution off the initial non-evolved sequence (eg. 
LDZ 1990; Sandage 1990a; Brocato et al. 1994, 1996; Clement \& 
Sheldon 1999a,b; Cacciari, Corwin, \& Carney 2004). 

The evidence for a period-amplitude-metallicity relation at 
various temperatures remains as it was set out in 1981 (S81a); it 
has been rediscovered by Carney, Storm, \& Jones [1992, their Eq. 
(16)], and by Alcock et al. (1998; Alcock 2000). 

It is useful to recall the strong debate in the 1990s 
concerning the observational data on period shifts. The extant 
ZAHB models did not predict such shifts when the predictions were 
made in a particular way, and the theoreticians, believing their 
models, questioned the interpretation of the observations on 
amplitudes and temperatures. The reasons were these.   

     The early pioneering models of the luminosity level of the 
ZAHB level for different metallicitites had been made by Sweigart 
and Gross (1976) followed by Sweigart, Renzini, \& Tornambe (1987, 
hereafter SRT), Bencivenni et al. (1991), and Dorman (1992). A 
more complete list of references is in SRT. Modern updates of 
these pioneering papers are, among others, those of Bono et al. 
(1997a), VandenBerg et al. (2000), and Catelan, Pritzl \& Smith 
(2004), each of which contain references to other recent models.

The period shifts with metallicity could be predicted from 
these models in the following way. The models give calculated 
tracks for the ZAHB in HR diagrams of Log L, log T$_e$ for different 
metallicities. They also give mass as a function of [Fe/H] along 
the tracks. Imposing a constant temperature line across the 
tracks, generally at log T$_e$ = 3.85, permitted log L and mass to 
be read along this fixed temperature line. A pulsation equation, 
P(L,M,T$_e$,[Fe/H]) that relates period, luminosity, and mass, is 
then used to predict how periods along the constant temperature 
line should vary with metallicity. 

These predictions gave a near zero slope (S93b, Fig. 6b) to 
a period-metallicity correlation rather than the value of 
dlogP/d([Fe/H]) = -0.12 derived from the observations. The effect 
of the increased luminosity on the period as the metallicity is 
decreased was nearly cancelled by the corresponding increased 
mass along the ZAHB in these models, shown explicitly by Simon 
and Clement (1993, the first and second unnumbered equations in 
their section 8). And even if the effect of luminosity evolution 
from the ZAHB was included as in Gratton, Tornambe, \& Ortolani 
(1986), and LDZ (1990, their Table 2), these models predicted a 
slope of the period-metallicity relation to be only 
dlogP/d([Fe/H]) = -0.05, but, again, {\it when read at constant
temperature}. This was less than half of what is observed.  

The solution of reconciling the model predictions with the 
observations, obvious now but not at the time, was to drop the 
condition of reading the models at constant temperature, but 
rather to let the temperature be cooler for more metal poor RRL 
at their higher luminosities. This solution had been proposed for 
RRc variables by Simon and Clement (1993) in their seminal paper 
on the discovery of methods to obtain luminosities, chemical 
abundances, masses, and temperature from various Fourier 
components of the light curves for RRc variables. They stated, 
based on their theoretical expectations, that ``~---it is not 
correct to formulate a period shift at constant temperature (eg. 
Sandage 1990b, ApJ 350, 631).''

The same solution of cooler temperature for the Oosterhoff 
II variables was proposed independently using the pulsation 
equation for RRab variables at the fundamental blue edge 
(S93a,b). It was shown there that a variation of temperature at 
the BFE at the rate of dlogT$_e$/d([Fe/H]) = -0.012 would largely 
reconcile the prediction from the models and the observations of 
period-shifts, if a mild evolution is also introduced (S93b, Fig. 
6a compared to 6b). 

The purpose of the present paper is to return to the 
problem, demonstrating more directly the need for a temperature 
variation with metallicity at the fundamental blue edge of the 
instability strip. In making the argument we reverse the 
dependent and independent variables in a pulsation equation by 
making T$_e$ the independent variable to be determined by assuming 
the observed dependencies of log P, log L, and log mass on 
metallicity, all of which are better known now than in S93a.   

The organization of the paper is as follows. Section 2 shows 
an updated period-metallicity correlation using the data in the 
catalog of 302 field RRL with measured metallicities by Layden 
(1994) on the scale of Zinn and West (1984). There are nearly 
three times the number of field stars in the Layden sample than 
in the catalog by Blanco (1992) that was used in S93a.     

Section 3 shows a similar period-metallicity distribution as 
in \S  2 for RRL in globular clusters from the updated catalog of 
cluster variables by Clement (2001) that merges the data from the 
third catalog of Sawyer Hogg (1973) with new data complete to 
2000 from the literature, adding about 30\% to the data base used 
in S93a.    

Section 4 sets out four modern calibrations of the absolute 
magnitude-metallicity relation for RRL by Caputo et al. (2000), 
Clementini et al. (2003), Catelan, Pritzl, \& Smith (2004), and 
McNamara et al. (2004), many of which emphasize the need for a 
non-linear M$_V$([Fe/H]) luminosity-metallicity relation both for 
the ZAHB and for the evolved HB. These four are a representative 
subset of many other current calibrations, reviewed by Cacciari 
and Clementini (2004). 

A new calibration of the mass-metallicity relation 
calculated by Bono et al. (1997a) is given in \S  5, and  
compared with that of Dorman (1992) that was used in S93a.
    
In \S 6 we combine the period-metallicity, M$_V$-metallicity, 
and mass-metallicity relations from \S\S 2-5 
with a pulsation equation to predict the temperature-([Fe/H])  
relation at the FBE. Temperatures are transformed to (B-V)$_o$ 
colors in \S 7 using an adopted color-metallicity-temperature 
calibration.
 
The predicted color-metallicity relation is compared in 
\S 8 with the observed (B-V)$_o$ colors for both field and cluster 
RRL. 

In \S 9 we turn the problem around by using the new 
correlations of period, temperature, and mass at the FBE to 
predict the RRL luminosity, again from the pulsation equation 
using a new temperature variation at the FBE that has been fitted 
empirically here to the observational data, rather than using the 
predicted variation in \S\S 6 \& 7. 

     Section 10 contains a discussion of the observed difference 
in the zero-point of the log period-([Fe/H]) envelope locus 
(Figs. 1-4) between field and cluster variables, and the presence 
or absence of the Oosterhoff period gap in the cluster and the 
field star samples, respectively, in terms of a proposed 
different formation history of the samples. 

     Eight research points made in this paper are summarized in 
\S  11. 

\section{THE DISTRIBUTION OF METALLICITY WITH PERIOD FOR THE 302 
         FIELD RR LYRAE VARIABLES IN LAYDEN'S SAMPLE}

In the 1993 study (S93a), I had used the correlation of mean 
period with metallicity for 110 field RR Lyraes for which Blanco 
(1992) had determined improved E(B-V) reddenings. He also had 
homogenized the existing [Fe/H] metallicity values based on 
Butler's (1975) calibration of the Preston $\Delta$S metallicity 
parameter. The Blanco-Butler [Fe/H] scale averages 0.28 dex more 
metal rich than the globular cluster scale of Zinn and West 
(1984). I had used the two different metallicity scales, one for 
the field variables and the other for the clusters, although I 
had transformed the Blanco-Butler scale to that of Zinn and West. 
Nevertheless, the field star log P-[Fe/H] diagrams were kept on 
the Blanco-Butler system, which, at times was inconvenient.    

This inconvenience is overcome here, and the sample of field 
stars has been increased by a factor of three by using the more 
recent data base of Layden (1994, 1995) which is on the Zinn-West 
metallicity scale. Layden measured metallicities for his complete 
sample of 302 RRab Lyrae field variables whose Galactic latitudes 
are more than $\pm 10^0$, and whose apparent magnitudes are brighter 
than $\langle V\rangle$ = 13.5. Layden's list is from the Fourth Edition of the 
General Catalog of Variable Stars (Kholopov 1985).  

The log P-[Fe/H] distribution for Layden's field star 
sample is shown in Figure 1. A linear envelope line is drawn by 
eye as the locus of the shortest period variables at a given 
[Fe/H] value. This would define the period at the fundamental 
blue edge of the instability strip if there is no ``hysteresis'' 
within in a possible ``either-or'' transition zone for evolution 
tracks between the fundamental and first overtone variables. The 
equation of the envelope line in Fig. 1 is
\begin{equation}
          \log P = -0.484 - 0.074 ([Fe/H]).  
\end{equation}

Editor:  Place Figure 1 here

The mid-point ridge line, shown as a dashed line in Fig. 1, 
is found by averaging the log P values in narrow intervals of 
[Fe/H]. The data are listed in Table 1. The equation of this line 
is,  
\begin{equation}
            \log P = -0.416 - 0.098 ([Fe/H]).
\end{equation}

Also listed in Table 1 are the mean (B-V)$_o$ colors within 
each metallicity bin, calculated from the photoelectric summary 
of individual colors listed in column 5 of Nikolov, Buchantsova, 
and Frolov (1984, hereafter the Sophia Catalog), using the 
reddening corrections for each star based on the absorptions 
(divided by 3) measured by Layden (1994). 

     The linear fit (the solid line) to the shortest period 
envelope data in Fig. 1 is good, but not perfect. A better fit is 
achieved for metallicities smaller than -1.8 by bending the 
envelope toward longer periods. The straight lines in Fig. 2 
have the equations, put by eye, of,   
\begin{equation}    
            \log P = - 0.484 - 0.074 ([Fe/H]),
\end{equation}

for [Fe/H]) more metal rich than -1.8, and two dashed lines whose 
equations are 
\begin{equation}
            \log P = - 0.610 - 0.144 ([Fe/H]), 
\end{equation}
and
\begin{equation}
              \log P = -0.680 - 0.183 (Fe/H])\eqnum{4'}
\end{equation}
for [Fe/H]) more metal poor than -1.8. Equation (3) is the same 
as Eq. (1) but its application stops for [Fe/H] smaller than      
-1.8.       

Editor:  Place Figure 2 here

     A still better fit is the continuous parabolic envelope over 
the entire metallicity range from zero to -2.4 shown in Fig. 3, 
whose equation is, 
\begin{equation}
 \log P = -0.452 + 0.033 ([Fe/H])^2,
\end{equation}
used previously for a different purpose (Sandage 2004, Fig. 5) 
using a subset of the present field star data.

Editor:  Place Figure 3

     These various envelope lines will be used in \S  6 for the 
predictions of the temperature at the FBE.    

\section{THE PERIOD DISTRIBUTIONS FOR DIFFERENT METALLICITIES 
                FOR RRL IN GLOBULAR CLUSTERS}

Although the number of stars (302) in Figs. 1-3 is 
sufficient to show the {\it continuous} variation of mean period with 
metallicity (rather than a dichotomy into two period groups), 
nevertheless it is not yet sufficient to define the position of 
the short period envelope at the low metallicity end for [Fe/H] $<$ 
-2 with precision. A larger sample of RR Lyrae variables is 
needed and is available in globular clusters. The earlier sample 
from the Third Catalog of cluster variables by Sawyer Hogg (1973) 
was used in S93a for this purpose.

     We have repeated the log period-metallicity diagram for the 
globular cluster variables in Fig. 4, increasing the cluster 
sample from 760 stars used by S93a to 919 here by adding the new 
variables listed in the 2001 updated catalog by Clemment (2001) 
cited earlier. The metallicity scale is that of Zinn and West, 
taken from the listing in the catalogs of Harris (1996).  The 
parabolic envelope of equation (5) is overlaid on the data to 
compare the cluster data with the field star data in \S 2 (Figs. 
1-3).  

Editor:  Place Figure 4 here

Two significant differences are evident by comparing Fig. 4 
for the cluster variables with Figs. 1-3 for the field variables. 

     (1). The gap in the [Fe/H] distribution (that is, in the 
vertical distribution of points) in Fig. 4 between [Fe/H] of -1.7 
and -2.0 for the cluster variables is absent in Figs. 1-3 for 
Layden's field variables. It is this gap that makes the 
Oosterhoff {\it dichotomy} into two discrete period groups so evident, 
partially masking a {\it continuous} variation of mean periods with 
metallicity. 

     (2). The parabolic envelope that fits the short period 
boundary of the distribution of periods for the field variables 
in Fig. 3 fails to fit the data for the globular cluster 
variables. The cluster variables have longer periods at the 
envelope edge by $\Delta$log P $\sim$ 0.029 dex compared with equation (5). 
The difference has a significance of about 4 sigma. 

There are several possibilities for explaining these two 
differences between field and cluster variables.     

     Point (1). The very few variables in the cluster sample in 
Fig. 4 between [Fe/H] = -1.7 and -2.0 is due, as is well known, 
to the details of the morphology of cluster horizontal branches 
through the instability strip in this metallicity range. Most 
clusters with intermediate metallicities with -1.3 $>$ [Fe/H] $>$  
-1.7 have horizontal branches well populated on both sides of the 
RR Lyrae instability strip, whereas clusters with [Fe/H] between 
-1.7 and -2.0 produce very few variables because the HB does not 
penetrate the instability strip. 

     Clearly, the absence of this Oosterhoff period gap in the 
field variables shows that the morphology of the horizontal 
branch of the field RRL {\it must} differ fundamentally from that in 
globular clusters in this metallicity range.  

The reasons are not presently understood, but a hint may be 
the variation of the HB morphology with metallicity of the highly 
unusual globular clusters NGC 6388 and NGC 6441 (Rich 1997 et 
al.; Pritzl et al. 2000, 2001) where the HB penetrates the 
instability strip despite its high metal abundance of              
[Fe/H] $\sim$ -0.5. This behavior is contrary to the horizontal 
branches of ``normal'' globular clusters such as 47 Tuc, and the 
clump HBs of high metallicity galactic clusters. Evidently, there 
is some parameter-variation (such as the value of the core He 
abundance, the core mass, variations in deep mixing, etc.) in 
these two highly abnormal globular clusters that causes the HB 
morphology to differ from that of most other of the clusters in 
the Galaxy. 

     Point (2). The observed period difference of 0.029 dex 
between cluster and field variables at the FBE can be 
accomplished in several ways from the pulsation Eq. (16) in \S 5. 
Among them is a luminosity difference of $\Delta$log L = 0.027 at 
constant temperature and mass, or a temperature difference of 
$\Delta$log T$_e$ = 0.007 at constant L and mass, or a mass difference of   
$\Delta$log mass = 0.034 at constant L and T$_e$, or combinations of each. 
It need only be recalled that Gratton (1998) has already 
discussed the possibility that there may be ``a real difference 
between the luminosity of the horizontal branch for clusters and 
the field''. 

     Point (3). A third possibility is that the metallicity 
scale due to Layden (1994) in Figs. 1-3 for the field variables 
is not the same as the used by Harris (1996) for the clusters. 
The difference seen between Figs. 1-3 and Fig. 4 would disappear 
if the field points in Figs.1-3 were moved downward by $\Delta$[Fe/H]   
$\sim$ 0.3 dex, or if the cluster data were made more metal poor by the 
same amount. But no systematic shifts by this amount are 
possible if the statements by both Layden and by Harris are 
correct that their scales are both tied to the scale of Zinn and 
West (1984).   

\section{ SELECTED RECENT ABSOLUTE MAGNITUDE CALIBRATIONS OF RR LYRAE 
              VARIABLES AS A FUNCTION OF [Fe/H] }

Many calibrations, both observational and theoretical, of RR 
Lyrae luminosities as functions of metallicity have been made 
since the linear calibration via the pulsation equation was made 
in S93a.                

     All of the second and third generation theoretical models 
for the position of the {\it zero-age} horizontal branches show that 
the relation between M$_{\rm bol}$ (and hence closely M$_V$) and [Fe/H] must 
be non linear, i.e. the branch loci of the ZAHB tracks are more 
closely stacked in luminosity at the low metallicity end of the 
distribution than at the high metallicity end, for equal 
intervals of [Fe/H], even as the low metallicity ZAHBs are 
brighter than those of high metallicity. Examples, among many, 
are the models of Lee, Demarque \& Zinn (1990); Castellani, 
Chieffi, \& Pulone (1991); Bencivenni et al. (1991); Dorman 
(1992); Caputo et al. (1993); Caloi, D'Antona \& Mazzitelli 
(1997); Salaris, Degl'Innocenti \& Weiss (1997); Cassisi et al. 
(1999); Demarque et al. (2000); VandenBerg et al. (2000, Figs. 2, 
3, and 20); Catelan, Pritzl, \& Smith (2004). A graphical summary 
showing several of the predicted M$_V$(HB)-[Fe/H] relations 
derived from these models is given by Cacciari \& Clementini 
(2004).     

     Observational and/or semi-theoretical calibrations that also 
show the need for a non-linear M$_V$([Fe/H]) relation are, among 
others, the studies by Caputo (1997); Gratton et al. (1997): De 
Santis \& Cassisi (1999, Fig. 15), Caputo et al. (2000), and 
McNamara et al. (2004).    
      
We use several of these calibrations in the pulsation 
equation to predict the temperature at the FBE of the 
instability strip in the next section, replacing the linear 
calibration derived in S93b of, 
\begin{equation}
            M_V =  0.94 + 0.30 ([Fe/H]).  
\end{equation}

We use the non-linear calibrations of Caputo et al. (2000), 
Catelan et al. (2004), and McNamara et al. (2004) from their tie-in 
to Delta Scuti stars, and the linear calibration of Clementini 
et al. (2003) from their calibration via the RR Lyrae stars in 
the LMC. The equations are assumed to be as follows.   

      We have approximated the graphical representation of the 
calibration of Caputo et al. (2000, their Fig. 2a) by a parabola, 
as, 
\begin{equation}
     M_V = 1.576 + 1.068 ([Fe/H]) + 0.242 (Fe/H])^2.
\end{equation}
Using M$_{\rm bol}$ = M$_V$ + BC, with BC = 0.06 +0.06 ([Fe/H]), and 
M$_{\rm bol}$(sun) = 4.75, it follows from (7) that   
\begin{equation}
    {\rm Log} L_{\rm bol} = 1.245 - 0.451 ([Fe/H])-0.097 ([Fe/H])^2.
\end{equation}              

The adopted bolometric correction here is the same as was 
used by Sandage and Cacciari (1990), justified there. The   
tables by Bell and Tripicco in Sandage et al. (1999, Table 6) 
show, of course, that the BC varies also with temperature and 
atmospheric turbulent velocity as well as with metallicity. 
Interpolation within the tables for turbulent velocity of          
5 km s$^{-1}$ at a color of (B-V)$^0$ = 0.24 (which is close to the blue 
edge) gives BC = 0.05 + 0.06([Fe/H]). Using a turbulent velocity 
of 1.7 km s$^{-1}$ gives BC = 0.06 +0.069 ([Fe/H]), again at (B-V)$^0$ = 
0.24. These justify our continued use of 0.06 + 0.06 ([Fe/H]) 
here.   

     The calibration equations from the theoretical models of the 
ZAHB by Catelan et al. (2004) using an oxygen enhancement of 
[alpha/Fe] = 0.3 and [Fe/H] = [M/H] - 0.213 give,        
\begin{equation}
    M_V = 1.179 + 0.548 (Fe/H]) + 0.108 ([Fe/H])^2,   
\end{equation}
from their Eq. (12) after some reduction to change [M/H] into 
[Fe/H]. This transforms in the same way as above to 
\begin{equation}
     \log L_{\rm bol} = 1.404 -0.243 ([Fe/H]) - 0.043 (Fe/H])^2.
\end{equation}

     The linear calibration of Clementini et al. (2003) using 
the observed RR Lyraes in the LMC and an adopted LMC modulus of 
(m-M) = 18.54 gives, 
\begin{equation}
          M_V = 0.845 + 0.214 ([Fe/H]),
\end{equation}
or,
\begin{equation}
          \log L_{\rm bol} = 1.538 - 0.110 (Fe/H]).
\end{equation}

Equations (11) and (12) refer to the mean (evolved) HB. To change 
to the level of the ZAHB we must make the equations fainter by 
about 0.1 mag in V, or 0.025 in log L (S90a).       

     The equations for the two-line calibration of McNamara et 
al. (2004) are; M$_V$ = 0.50 independent of [Fe/H] for metallicities 
more metal poor than -1.5, and M$_V$ = 1.13 + 0.42 ([Fe/H]) for 
[Fe/H] between -0.5 and -1.5. 

     These transform to 
\begin{equation}
         \log L_{\rm bol} = 1.676 - 0.024 ([Fe/H]),
\end{equation}
for [Fe/H] $<$ -1.5, and, 
\begin{equation}
         {\rm Log} L_{\rm bol} = 1.424 - 0.192 ([Fe/H]),
\end{equation}
for [Fe/H]) $>$ -1.5.

\section{THE DEPENDENCE OF RR LYRAE MASS ON METALLICITY }

To make the calculation of T$_e$(FBE) from the pulsation 
equation, we also need the mass as a function of metallicity. In 
S93b, I adopted the ZAHB calculation of RR Lyrae mass at log 
T$_e$ = 3.85 by Dorman (1992) as log mass = -0.288 - 0.059 ([Fe/H]), 
where he had used enhanced oxygen abundances and modern opacity 
tables. More recently this theoretical calculation of RR Lyrae 
masses has been verified and extended by Bono et al. (1997a), 
again using enhanced oxygen abundances and the latest opacity 
tables cited by them. The resulting masses on the ZAHB at log T = 
3.85 are listed in Table 2 of Caputo et al. (2000), and are 
adopted here. The resulting linear mass equation over the 
relevant metallicity range is, 
\begin{equation}
          {\rm Log mass} = -0.283-0.066 ([Fe/H]),
\end{equation}
which we adopt. This deviates by not more than 0.02 dex in mass 
from Dorman (1992) over the metallicity range from 0 to -2.5. 

     Confirmation that the mass of metal poor RR Lyraes is larger 
than that for metal rich variables is available from analyses of 
the observations of double mode variables via the Peterson (1973) 
diagram and the subsequent calculations by Peterson (1978, 1979), 
Cox et al. (1980), Cox et al. (1983), and Cox (1987) and 
references therein. The calculations by Kovacs et al. (1992) that 
show the sensitivity of the results to various chemical 
composition mixtures are illustrated in Fig. 4 of S93b. 

     The masses from Eq. (15) are for the ZAHB at log T$_e$ = 3.85. 
They will not be precise even at this fixed temperature if 
evolution brings stars from different parts of the HB into the 
instability strip. We neglect the detail in this reconnaissance 
study; it does not change the sense of the argument that follows 
in Figs. 5 \& 6 of the next section.  

\section{PREDICTED VARIATION OF TEMPERATURE WITH METALLICITY AT THE 
                  FUNDAMENTAL BLUE EDGE }

Many new versions of the pulsation equation have been 
calculated since the pioneering paper by van Albada and Baker 
(1973) (eg. Chiosi et al. 1992; Simon \& Clement 1993; Bono et al. 
1997b with earlier references), but each give nearly identical 
dependencies of period on temperature, luminosity, and mass. We 
continue to use the van Albada and Baker formulation here.   

     Their equation, transposed to make temperature the 
independent variable, is        
\begin{equation}
  \log T_e= 0.241 \log L - 0.287 \log P - 0.196 \log M + 3.304.
\end{equation}

If the dependencies of L, P, and mass on metallicity have 
been expressed as relations at the FBE of the instability strip, 
either on the ZAHB or on the mean HB as evolved from the ZAHB, 
then Eq. (16) will give the temperature at that edge. We 
make the assumption here that the HB is truly horizontal in V 
magnitudes and, therefore, that equations (8), (10), (12), 
(13), and (14) also define the bolometric luminosity at the blue 
fundamental edge as well as at the mean strip position to which 
many of the luminosity calibrations of the aforesaid equations 
apply. This is a fully adequate approximation for the restricted  
purpose of this paper which is only to show the {\it need} to invoke a 
variable temperature with metallicity at the FBE of the strip 
rather than to finalize a definitive determination of it. 

     Indeed, uncertainties presently exist in the temperature-
color calibration (\S 7) at the level of 0.02 mag in B-V. 
Similar uncertainties also exist at the 0.02 mag level in the 
observational data for the B-V color at the FBE edge due to 
uncertainties in the reddening. In addition, theoretical 
uncertainties exist in the constants of the pulsation equation 
including its zero point [i.e. 3.304 in Eq. (16)]. These all 
conspire against a definitive ``final'' calculation of T$_e$(BFE). 
Nevertheless, all log L calibrations to date, and all P-([Fe/H]) 
displays such as in Figs. 1-4, require the same qualitative 
variation of T$_e$ and color of the fundamental blue edge with 
metallicity which is the purpose of this paper, which we now 
demonstrate. 

      To illustrate the nature of the 30 solutions for T$_e$ in 
the completely filled log P([Fe/H]) and log L([Fe/H]) parameter 
space, it is sufficient to choose only a subset of the P([Fe/H]) 
and L([Fe/H]) combinations. Here the subsets are divided into two 
groups. In the first we keep the luminosity calibration fixed 
using Eq. (10) by Catelan et al. (2004), and use with it the 
four different log P-metallicity envelopes from Figs. 1-3. These 
loci are (a) the log P([Fe/H]) envelope in Eq. (1) as in Fig. 1; 
(b) the three-line envelopes of Eqs. (3), (4) and (4') in Fig. 2; 
and (c) the parabolic envelope of Eq. (5) in Fig. 3. 

Inserting the Catelan et al. (2004) log L([Fe/H]) relation 
from equation (10), the Bono et al. (1997a) mass equation from 
Eq. (15), and the log P([Fe/H]) relations from Eqs. (1), (3), 
(4), (4') and (5) into the pulsation equation (16) gives the run 
of predicted log T$_e$ with [Fe/H] shown in Fig. 5. The relation 
using the midpoint correlation of [Fe/H] with period [Eq. (2)] is 
also shown.   

Editor:  Place Figure 5 here

     As an example, the predicted log T$_e$ relations at the FBE 
using the parabolic envelope line of Eq. (5), the Bono mass 
Eq. (15), and the Catelan et al. luminosity calibration of Eq. 
(10) is,   
\begin{equation}
\log T_e (FBE) = 3.827 -0.046 ([Fe/H) - 0.019 ([Fe/H])^2,
\end{equation}
shown in Fig. 5 as the dashed line marked number 3. 

     Alternatively, by keeping the envelope log P([Fe/H]) 
relation fixed using Eq. (5) for the parabolic fit, and using 
five different Log L(Fe/H]) calibrations give the predicted run 
of log T$_e$ shown in Fig. 6.  

Editor:  Place Figure 6 here

The characteristics of all the solutions in both diagrams is 
that the predicted temperature at the fundamental blue edge 
varies with [Fe/H], first becoming hotter as [Fe/H] weakens from 
0 to within the range of -1.2 to -1.5, and then cools again for 
still decreasing metal abundance. At no metallicity is it 
constant at the FBE with varying [Fe/H], although its variation 
is less than $\Delta$log T$_e$ = 0.01 between [Fe/H] of -0.5 and -2.0 
along any given line of fixed luminosity calibration. 

\section{TRANSFORMATION OF THE TEMPERATURE VARIATION AT THE FBE 
                      TO B-V COLORS} 

     A number of new calibrations of how B-V color varies with 
log T$_e$, metallicity, surface gravity, and atmospheric turbulent 
velocity have been made since the relation, taken from an 
unpublished calibration of Bell, as quoted by Butler et al. 
(1978, their Fig. 11), was used by S93b (Eq. 4 there). Six of 
these modern calibrations have been compared by Cacciari, Corwin, 
\& Carney (2004, their Fig. 7). The total differences between the 
various calibrations in the predicted B-V colors at a given 
temperature near 7000 K (log T = 3.845) is large at 0.08 mag. Or 
read the other way, at a color of B-V = 0.28, the temperatures 
range between 6950 K (log T$_e$ = 3.842) to 6400 K (log T = 3.806) 
for the calibrations that are compared by Cacciari et al. (2004). 

The hottest calibrations for a given color are those of 
Sandage, Bell, and Tripicco (1999, hereafter SBT) and Castelli 
(1999). The coolest is that of Montegriffio et al. (1998). 
Intermediate are the calibrations of Carney, Storm, \& Jones 
(1992), and Sekiguchi \& Fukugita (2000). The bulk of the 
comparisons show a smaller spread of 0.04 mag in B-V between the 
calibrations of SBT, SF, and CSJ.
       
     We adopt the intermediate calibration of CSJ (1992, their 
Eq. 13), which, in the temperature range of interest (log T$_e$ 
between 3.85 and 3.80), can be put with sufficient accuracy into 
the logarithmic form rather than as function of 5040/T as they 
give it, to be, 
\begin{equation}
     (B-V) = -2.632 \log T_e + 0.038 ([Fe/H]) + 10.423,
\end{equation}
in the interval 0.20 $<$ B-V $<$ 0.30.  

Comparison between the SBT and the SF scales read at the 
fixed color of B-V = 0.24 at [Fe/H/] = -1.5 (near the mid range 
of the relevant parameter space), give log T$_e$ = 3.848 (T$_e$ = 7046 
K) for CSJ from eq. (18), 3.857 (T$_e$ = 7194 K) for SBT, and 3.852 
(T$_e$ = 7112 K) from SF. Hence, the temperatures on the three 
scales differ by $\Delta$log T$_e$ = 0.009 at B-V = 0.24. 

     The coefficient of [Fe/H] in Eq. (18) is, of course, a 
strong function of temperature, increasing due to Fraunhofer 
blanketing and atmospheric backwarming with decreasing 
temperature. The value of 0.038 in Eq. (18) is a compromise 
between 0.026 at B-V = 0.24 for the SBT models (interpolating in 
their Table 6) and 0.047 in the SF calibration, a value that is 
surely too large at the hot temperature corresponding to B-V = 
0.24. 

      Using Eq. (18) we transform the T$_e$ at the FBE for each of 
the equations that generate the family of curves in Figs. 5 and 6 
into the predicted run of B-V colors with metallicity at the FBE. 

     To illustrate one equation of the family we insert the 
base-line parabolic equation for the log P([Fe/H]) envelope 
relation of Eq. (5) and the Catelan et al. luminosity calibration 
of Eq. (10) into Eq. (18) to obtain the predicted run of B-V color 
at the FBE for that combination as, 
\begin{equation}
   (B-V)_{\rm FBE} = 0.35 + 0.159 ([Fe/H]) + 0.050 (Fe/H])^2,
\end{equation}
which is the dashed line 3 in Fig. 7.

     The complete sets of the families from Figs. 5 and 6, but 
now in B-V, are shown in Figs. 7 and 8. 

Editor:  Place Figures 7 and 8 here

     All curves in Figs. 7 and 8 have the same character. Each 
predicts a bluing as [Fe/H] decreases from 0 to about              
-1.5, after which the predicted color becomes redder. Over the 
relevant range of metallicities from [Fe/H] = -0.9 to -2.2, which 
contains most of the Galactic globular clusters, the color 
change is only $\Delta$(B-V) $\sim$  0.02 mag. 

     We emphasize that the zero points of the B-V values in 
Figs. 7 and 8 have not been adjusted to conform with 
observational data. They are the calculated from the theoretical 
zero points in the several relevant equations. Given the range of 
errors in each of these equations, it is remarkable that the 
agreement of the predicted color variation with the observations, 
to be discussed in the next section, is so good.

\section{ COMPARISON OF PREDICTED AND OBSERVED B-V COLORS OF FIELD AND 
        CLUSTER VARIABLES AT THE FUNDAMENTAL BLUE EDGE}

\subsection{Which definition of mean color to use?}

Which definition of ``mean color'', averaged over the light 
curve, best approximates the color of an equivalent static star? 
Three definitions are used in the literature, defined by the 
following operational procedures. (1) Change B and V magnitudes 
into intensity units and average over the light curve, 
calculating mean values of $\langle B\rangle_{\rm int}$ and $\langle
V\rangle_{\rm int}$, changed back to 
magnitude units, and then subtract to form the color $\langle B
\rangle_{\rm int}$ - 
$\langle V\rangle_{\rm int}$. (2) Change the color curve in B-V from magnitude to 
intensity units and average over the intensity color curve, 
changed back to magnitude units to form $\langle B-V\rangle_{\rm int}$ 
(3) Keep the 
color curve in magnitude units and integrate over the cycle to 
obtain $(B-V)_{\rm mag}$. 

     These procedures give colors that can differ by as much as 
0.05 mag for highly asymmetric light and color curves of high 
amplitude. Which is correct to give an approximation to the color 
of a ``static star'' with the same average energy output? 

     In a semi-analytical treatment, Preston (1961) showed that 
the best estimate of the color of such a static star is 
$(B-V)_{\rm mag}$. 
I reached the same conclusion (S90a) in a semi-empirical 
discussion based on the calculated and observed slopes of the 
period-color relation calculated from the pulsation equation, 
choosing a color equivalent definition that made the predicted 
and observed slopes the same. Corwin and Carney (2001) also 
concluded that $(B-V)_{\rm mag}$ was the 
appropriate color to use, 
reversing a previous discussion by Carney, Storm, and Jones 
(1992) to the contrary. A definitive theoretical discussion by 
Bono, Caputo, \& Stellingwerf (1995) showed that $(B-V)_{\rm mag}$ is 
within 0.01 mag of $(B-V)_{\rm static}$ for all blue amplitudes smaller 
than 1.6 mag. A table of differences between the three types of 
color definitions and $(B-V)_{\rm static}$, which they calculated, is 
given by them.    

     We use $(B-V)_{\rm mag}$ in the remainder of the paper. We have 
converted other listings in the literature to $(B-V)_{\rm mag}$
where they 
differ, either by using Table 4 of Bono et al. (1995), or using 
the $\Delta$C(A) correction formulation in S90a. A detailed reading of 
the literature is often required to determine which mean color 
definition has been used by particular authors.     

\subsection{Observed $(B-V)^0_{\rm mag}$
for a sample of field variables}

We have attempted to assemble a data base of reddening-
corrected $(B-V)^0$
colors of RR Lyrae variables accurate at the 
0.02 mag level for both field and clusters variables. If the 
observational data are from a variety of observers, they must be 
reduced to a common photometric standard system, taking out 
systematic zero-point differences between the sets. In addition, 
a uniform definition of what kind of mean color, as just 
discussed, must be used. More uncertain are the corrections for 
E(B-V) reddening, that themselves are generally be accurate only 
to within 0.02 mag. Hence, detection of the relatively small 
color variations with [Fe/H] predicted in Figs. 7 and 8 is near 
the limit of accuracy of the data now available.    

    A large list of observed photoelectric colors for field RR 
Lyrae variables is in the catalog by Nikolov, Buchantsova, \& 
Frolov (1984), hereafter referred as the Sofia Catalog. Nine 
photoelectric data sets from the literature were compared and 
reduced to a common system, producing a highly homogeneous list 
of $(B-V)_{\rm mag}$
colors for 210 variables in that catalog (their Table 2, 
column 2). The photoelectric data, so reduced, are from Fitch et 
al. (1966), Sturch (1966), Clube et al. (1969), Paczynski 
(1965a,b, 1966), Stepien (1972), Epstein (1969), Lub (1979), 
Preston \& Paczynski (1964), and Kinman (1961).   

I have used the listings in the Sophia catalog and corrected 
them for reddening by using the homogeneous set of absorption 
values in V measured by Layden (1994), using E(B-V) = A(V)/3.0 
for all stars in common with the Sophia catalog and the Layden 
list. In this way, $(B-V)_o$ colors were obtained for 142 field RR 
Lyraes in the Layden sample, plus 10 additional from Blanco's 
study discussed in S93a but not in Layden's list, and an 
additional 8 from Lub that are not in Layden but reduced to his 
absorption system by cross checking the Lub list as constructed 
in S90a with the Layden list. A total of 160 field variables with 
$(B-V)_o$ colors are available in this way. The color values are not 
listed here because they can be readily obtained from the same 
original sources. 

Editor:  Place Figure 9 here

     The colors for these field variables are plotted in Fig. 9 
vs. Layden's [Fe/H] metallicities. As emphasized earlier, the 
[Fe/H] values are on the system of Zinn and West (1984). The line 
is equation (19), which is the prediction of T$_e$ using the 
pulsation equation (16) and the transformation to B-V via Eq. 
(18). It is the predicted color-metallicity locus at the FBE 
using the Catelan et al. (2004) luminosity calibration of the FBE 
in Eq. (10) and the parabolic envelope locus to the period-
metallicity relation of Eq. (5) in the pulsation equation for T$_e$, 
transformed to B-V via Eq. (18). The drawn line is repeated from 
Figs. 7 and 8, which is one of the curves in the family shown 
there.         

No adjustment of the {\it zero point} of the theoretical equation 
(19) to fit the observations has been made in Fig. 9. Hence, as 
stated earlier, the generally good agreement of this calculated 
curve with the observations of the bluest color at a given 
metallicity is remarkable because the various zero points in the 
theoretical and the observed equations used for the prediction 
have been left as they had been previously calculated.
 
The predicted T$_e$([Fe/H]) line in Fig. 9 fits the 
observations tolerably well except for metallicities more metal 
poor than [Fe/H] = -2.0 where it does not bend sufficiently 
toward redder colors. A better representation is derived in the 
next section (Fig. 11 there) where the color data for variables 
in globular clusters are combined with these field star colors. 

\subsection{The Observed $(B-V)^0_{\rm mag}$ 
Colors for Globular Cluster RRab 
       Lyrae Variables From the Literature CCD Photometry}

The globular cluster color data are set out in Fig. 10. They  
have been taken from the literature, either from modern CCD 
photometry, or from reliable pg photometry. The literature 
citations for the 21 clusters are listed in Table 2. Most data 
have been transformed to the $(B-V)_{\rm mag}$ 
mean color system by Table 
4 of Bono et al. (1995). However, if the original data were 
originally listed as $(B-V)_{\rm static}$, 
they were kept at that because, 
according to Bono et al., the ``static star'' color is within 0.01 
mag of $(B-V)_{\rm mag}$ for all amplitudes of interest. 

Editor:  Place Figure 10 here.

     The E(B-V) reddening values determined by the original 
authors have been replaced by the homogeneous values listed by 
Harris (1996), except for NGC 4590 where the value proposed by 
Walker (1994) is obviously more correct. 

Editor:  Place Figure 11 here. 

The data in Figs. 9 and 10 are combined in Fig. 11. Although 
there are outriders to the bulk of the distribution at blue 
colors, the trend is for the clusters with [Fe/H] between -1.0 
and -1.7 to be bluer at the short period edge than either for the 
more metal rich or the more metal poor variables, similar to the 
solid curves in Figs. 9 and 10. A stronger bend toward the red is 
drawn as a new curve fitted empirically to the data for [Fe/H] $<$   
-2.0 in Fig. 11. It is a slightly better fit to the combined data 
than the curve that is drawn in Fig. 9 and 10 using Eq. (19). 
The equation of this empirical curve is,   
\begin{equation}
    B-V = 0.351 + 0.172 ([Fe/H]) + 0.061 ([Fe/H])^2 
\end{equation}
This can be transformed to temperature via Eq. (18) to give 
\begin{equation}
   log T_e = 3.827 - 0.051 ([Fe/H]) -0.023 ([Fe/H])^2
\end{equation}
We aver that this applies at the FBE because we have assumed that 
the blue envelope loci in Figs. 3 and 10 (solid line) refer to 
that edge. 

\section{ THE PROBLEM TURNED AROUND TO PRODUCE A NEW LUMINOSITY 
                        CALIBRATION }

We now repeat the calculation of the luminosity calibration 
with metallicity made in S93a but here we use the new 
correlations between period, temperature, and mass relations with 
metallicity set out in the previous sections. The method is to 
insert the parabolic Eq. (5) for the period-metallicity locus at 
the FBE, the new semi-empirical equation for the temperature 
at that edge from Eq. (21), Fig. 11, and the mass-metallicity 
relation from Eq. (15) into the pulsation equation, permitting the 
calculation of the luminosity.  
             
     Using Eqs. (21), (5), and (15) in Eq. (16) gives,
\begin{equation}
 log L_{\rm bol} = 1.401 - 0.264 (Fe/H]) - 0.056 ([Fe/H])^2.
\end{equation}
As before, using M$_{\rm bol}$ = -2.5 log L + 4.75 and BC = 0.06 + 0.06 
([Fe/H]) in M$_V$ = M$_{\rm bol}$ - BC gives, from Eq. (22),
\begin{equation}
     M_V = 1.187 + 0.600 ([Fe/H]) + 0.140 (Fe/H])^2,
\end{equation}
over the metallicity range of [Fe/H] $\sim$  -0.5 to -2.3. This 
calibration updates the linear calibration of Eq. (6) given in 
S93b, made by the same method, but using the new assumptions 
here that relate period, mass, and temperature variations with 
metallicity.   

     As already said several times, the zero point in equation 
(23) is the theoretical value based on the zero points adopted in 
each of the pulsation, mass, and color-temperature equations. 
Because we cannot guarantee that these combined zero points are 
precise, we change the theoretical zero point in Eq. (23) to 
conform with the independent data on M$_V$ (RR) determined 
empirically. Clementini et al. (2003) have measured the mean 
apparent magnitude of RRLs in the LMC as a function of [Fe/H], 
which can be changed to absolute magnitude by assuming a distance 
modulus of the LMC determined by other means. They adopt (m-M)$_o$ = 
18.54 based on an adopted P-L relation of the LMC type I long 
period Cepheids. However, Tammann, Sandage \& Reindl (2003) argued 
that the Cepheid P-L relation differs between the Galaxy and the 
LMC, and they fixed the distance modulus the LMC also at (m - M)$_o$ 
= 18.54 by means other then the long period Cepheids. 

     The data of Clementini et al. give $\langle V\rangle$ = 19.064 for the RR 
Lyraes at [Fe/H] = -1.5 in the LMC, which, with the assumed 
modulus of 18.54, give M$_V$ = 0.524. This differs by 0.078 mag from 
the zero point in Eq. (23) of M$_V$ = 0.602 for this metallicity. 
Hence, we must make the zero point in Eq. (23) brighter by 0.078 
mag to put it on the observed scale of the Clementini et al.  

     But the Clementini data refer to the average RR Lyrae 
apparent magnitude which already accounts for the average 
luminosity evolution from the ZAHB. A difference of about 0.1 mag 
is expected between the ZAHB and the average evolved HB 
level (S93b). Remarkably, this is near the 0.078 mag difference 
just quoted. It continues as a surprise that the calculated and 
observed zero points, corrected for evolution, are so nearly the 
same at the 0.10-0.078 = 0.02 mag level. Again, 
this can only mean that the zero points of 
the various relations that go into Eq. (23) are themselves fully 
compatible with the various observations that comprise them.   

     Therefore, our final calibration of the ZAHB using the 
theoreticians route via the pulsation equation, but zero-pointed 
through the LMC as in Eq. (23) and made brighter by 0.078 mag, 
is, 
\begin{equation}
    M_V = 1.109 + 0.600 ([Fe/H])  +0.140 ([Fe/H])^2,
\end{equation}
valid for the average evolved HB at the FBE.

\section{IS THERE A DIFFERENCE BETWEEN FIELD RR LYRAES AND THOSE 
                    IN CLUSTERS?}

It remains to discuss the significant of the difference  
between Layden's field star sample and the RRLs in clusters in 
the absence of the Oosterhoff gap in Fig. 1 and its presence in 
Fig. 4, and, in addition, in the shift of the short period 
envelope lines between of Figs. 1 and 4 in the period-metallicity 
correlations. Both differences were set out earlier 
in \S 3. 

The absence of a gap (i.e. the presence of a continuum) in 
particular Galaxian samples of field RRLs had already been 
seen in the original period-metallicity sample of Preston (1959, 
his Fig. 4). It is also highly manifest in the much larger sample 
of Layden (1994, 1995) in Figs. 1-3 here. The gap is also absent 
in recent discovery surveys of faint RRL in high latitudes (eg. 
Vivas \& Zinn 2002, 2004), A summary of the relevant evidence is 
by Catelan (2004)\footnote{However, contrary evidence exists showing the presence of 
the gap in the faint field RRL samples of Suntzeff, Kinman, 
and Kraft [1991], in contrast to Figs. 1-3 and to the results of 
Vivas \& Zinn (2002, 2004). Catelan (2004) remarks that ``the 
reason for the discrepancy between the two studies is unclear at 
present.''}    

Even more to the point, for some samples of the dwarf 
spheroidal companions to the Galaxy the consequence of the 
intermediate value of the mean period, 
$\langle P\rangle$, is that there is no Oosterhoff 
separation into a {\it dichotomy} for different metallicitues (Bono, 
Caputo, \& Stellingwerf 1994; Mackey and Gilmore 2003 ; Mateo 
1996; Pritzl et al. 2002; Cseresnjes 2001; Clementini et al. 
(2004); and Dall'Ora et al. (2003). The result is that the mean 
periods of the RRL are {\it intermediate} between the two Oosterhoff 
groups in the Galaxy (Catalan 2004, Table 1; Siegel \& 
Majewski 2000, their Fig. 6). Why? 

     The general morphology-metallicity progression for 
horizontal branches in the globular clusters in the Galaxy is 
this. The HBs in the high metallicity clusters for [Fe/H] $\gtrsim - 0.7$ 
do not penetrate the instability strip because they are 
redder than the strip color boundaries. Hence, such clusters 
contain few if any RR Lyrae variables. The prototype example is 
47 Tuc. 

 A counter example in the opposite sense is NGC 7006 which 
has an abnormally red segment to its HB for its metallicity 
(Sandage \& Wildey 1967), yet it contains many RRLs. Also in an 
opposite sense (abnormally blue HB for its metallicity) is M13 
where the branch misses the instability strip altogether  
because of its excessive blueness, yet its 
metallicity is nearly identical with that of M3 which contains 
many variables (cf. Cho et al. 2005 for a modern comparison 
of M3 and M13).  

Hence, the cluster morphology-metallicity relation that is 
responsible for the Oosterhoff gap in the Galaxy is not perfectly 
consistent. The most extreme examples of this second parameter 
effect are the metal rich ([Fe/H] $\sim$ -0.50) clusters NGC 6388 and 
NGC 6441 which have abnormally long periods for their amplitudes 
(Pritzl et al. 2000, 2001), yet both have abnormally blue 
extended horizontal branches never before seen in other high 
metallicity clusters (Rich et al. 1997). The globular cluster M2 
is a less extreme example (Lee \& Carney 1999b) with much longer 
periods for a given amplitude. Also, similar to M2, the second- 
parameter custer NGC 5986 (Alves, Bond, \& Onken 2001) has a 
period-amplitude diagram with abnormally long periods compared 
with even Oosterhoff II clusters, although the metal abundance is 
only -1.58 on the Zinn \& West system. Its HB is abnormally blue, 
as in M13 (van den Bergh 1967), for this metallicity. This is 
opposite the sense of the effect in NGC 7006. 

     Although not yet conclusively proved, the reasons for such 
behavior is currently suspected to be different chemical 
composition effects that control the details of the HB morphology 
(eg. variations in the helium core mass, differences in the mass 
loss from the AGB to the HB, differences in the deep mixing 
etc.). If so, the difference between the field and the cluster RR 
Lyraes (Figs. 1 and 4) would suggest that these field variables 
have had a different history in their chemical evolution than 
those in clusters. Yes, but why? 

Suppose that the bulk of the field RRLs have come from the 
disintegration of dwarf spheroidals that once were companions to 
the Galaxy but have since been accreated and disrupted. This is 
almost certainly proved now by the discovery of the discrete RRL 
streams in the Sloan survey for high latitude variables (Ivezic 
et al. 2000; Yanny et al. 2000; Newberg et al. 2002), and the 
similar discovery by Vivas et al. (2001 cf. also Vivas \& Zinn, 
2002, 2004) from the Venezuela QUEST survey. These discrete 
streams, widely assumed to be debris from original dwarf 
spheroidal companions to the Galaxy, now disrupted, are the 
generalization of the many high velocity moving groups discovered 
by Eggen (1977), the first of which was the Grombridge 1830 group 
that contains RR Lyrae itself (Eggen \& Sandage 1959). The Sloan 
survey, and other evidence before it (eg. Freeman 1987; Majewski 
1993; Lynden-Bell \& Lynden Bell 1995; Freeman \& Bland-Hawthorn 
2002), showed that the Eggen high velocity moving groups are 
real. The disruption of dwarf spheroidal Galaxian companions is 
seen directly today in the Sagittarius disintegrating dwarf 
(Ibata, Gilmore, \& Irwin 1994, 1995) as it is being torn apart by 
tidal interaction with the Galaxy.       
     
It is known that the chemical evolution of the individual 
dwarf spheroidals differ, depending on their individaual 
luminosity, i.e. mass (eg. Aaronson 1986; Skillman et al. 1989, 
Caldwell et al. 1998; Mateo 1998, his Fig. 7 for a summary). 
Hence, the HB morphology of the products of the disruption (among 
which are the halo RRLs), should also differ by second-parameter 
effects. In particular, as a group, they should differ from the 
``normal'' HB morphology of Galactic globular clusters where the 
chemical evolution has gone further toward ``completion'' because 
the much higher mass of the Galaxy means stronger gravity and 
therefore the chemical products of AGB nucleosynthesis can be 
retained, unlike the situation in the dwarf spheroidals where 
supernovae winds can cause mass loss from the parent dwarf. 

      Because of the variable ability of the parent galaxies to 
retain the products of the chemical build up, depending on the 
mass, the chemical evolution of the parent dwarf spheroidals has 
differed from that history for most of the globular clusters in 
the Galaxy that have the normal Oosterhoff gap. Hence, the 
details of the Oosterhoff effect can be expected to differ.   

     If so, one can no longer support the canonical idea that the 
field RR Lyraes have come from the general Galactic globular 
cluster population, but rather from the previous dwarf spheroidal 
companions with varying HB morphologies. 

     The hypothesis to be tested is this. The majority of the 
globular clusters in the inner Galactic halo have a common origin 
that coincided with the origin of the bulk of the Galaxy itself 
(van den Bergh \& Mackey 2004). The evidence supports the idea 
that this process is a more or less coherent collapse, albeit 
with noise, similar to that of Eggen et al. (1962, ELS; Sandage 
1990c) for the lower halo, the bulge, and thin, and thick disks. 
On the other hand, as argued above, many of the halo field 
variables have come from the ``outside'' in a type of Searle-Zinn 
(1978) bottom-up hierarchical fragmentary build up in some form 
of the cold dark matter scenario for origins. 

     However, the most telling argument against the bottom-up 
build up from fragments of the {\it bulk} of the Galaxy, and other 
similar high mass galaxies, is the recent discovery of normal E 
galaxies at the very large redshifts of z $>$ 2 (McCarthy 2004) 
which would not exist in the hierarchical build-up from small 
fragments. Rather, there existence is the natural consequence of 
an ELS-type collapse on time scale short relative to the Hubble 
time. Said differently, normal E galaxies undergoing passive 
evolution, but otherwise fully formed, cannot be created by slow 
hierarchical build up on the cold dark matter scenario, but must 
be formed very early, very rapidly, all components of a 
protogalxy having a negative total energy. This is direct ELS. 

     A test of the different-origins hypothesis for the field 
and the cluster variables in the Galaxy is to map the morphology-
metallicity relation of the HB of the dwarf spheroidals and to 
match the differences in the period-metallicity envelope, for 
example, with such second parameter indicators as the period-
amplitude relation, the period-rise tine correlation, and the 
period-color relations with the Oosterhoff period shifts as 
functions of morphology and metallicity. Such a study is beyond 
the scope of this paper but is in progress.  

\section{SUMMARY}

There are eight principle research points in this paper. 

     1. The best fitting envelope of the period-metallicity 
correlation (Fig. 1) at the shortest periods for the field RR 
Lyraes in the Layden (1994) sample is non-linear. A parabolic 
relation [Eq. (5)] more adequately fits the data at low metallicity 
than the linear envelope adopted earlier (S93a).  

     2. The shape of the envelope in item 1 is also a good fit to 
the period-metallicity relation for RRL in globular clusters 
(Fig. 4), except that it is displaced in zero point toward longer 
periods by $\Delta$log P = 0.029 $\pm$ 0.007. This is another piece of 
evidence, joining several already in the literature, that the 
field and cluster variables may have different origins (points 7 
and 8 below).   

     3. The equations for various period-metallicity envelopes in 
items 1 and 2 are inserted into the pulsation equation, together 
with a series of new calibrations of the absolute magnitude of RR 
Lyrae variables, and with an updated calibration of RR Lyrae mass 
with metallicity, to predict a family of temperature variations 
at the fundamental blue edge as function of metallicity. All 
combinations of the input assumptions show that T$_e$ at the FBE 
varies with metallicity, changing by $\Delta$log T$_e$ = 0.04 over the 
range of ([Fe/H]) between 0 and -2.5. It first becomes hotter as 
[Fe/H] changes from 0 to -1.5, and then cooler for as the 
metallicity decreases further (Figs. 5 \& 6).      

     4. The predicted FBE temperatures are transformed into B-V 
color (Figs. 7 \& 8) by the adopted temperature-color relation of 
Eq. (18). The color variation is predicted to be at the level of 
0.03 mag in (B-V) for [Fe/H] between -1.0 and -2.5 for any given 
choice of the absolute magnitude calibration and of the variation 
of the period-metallicity locus at the FBE.  

    5. Comparison of this predicted color variation for both 
field and cluster variables (Figs. 9 \& 10) shows general 
agreement in the shape of the variation and in the color zero 
point. The agreement of this theoretical color zero point with 
the observations is remarkable because the calculations are made 
adopting the zero points in the pulsation equation [Eq. (16)], the 
mass equation (15), the color-temperature transformation Eq. 
(18), and the luminosity calibration equations, each of which 
have a range of uncertainty. 

     6. An iteration of the predicted envelope locus for the 
color-metallicity correlations of Figs. 9 and 10 to better fit 
the color-[Fe/H] data for metallicities smaller than -2.0 gives 
the semi-empirical locus of Eq. (20), shown in Fig. 11. This 
envelope, now made to better fit the observations, is used to 
make the calculation via the pulsation backward to obtain an 
improved absolute magnitude calibration of Eq. (23) which refers 
to the age zero HB. This equation is then re-zero pointed to 
refer to the average evolved HB by adopting the Clementini et al. 
(2003) LMC observation that $\langle V\rangle_o$ = 19.064 at [Fe/H] = -1.5 for 
the LMC RRLs and using the modulus of the LMC as (m-M)$_o$ = 18.54 
(Tammann et al. 2003) which does not use the LMC classical 
Cepheids. This gives a new calibration of the average M$_V$ of the 
evolved HB of Galactic globular clusters [(Eq. (24)].   

     7. Returning to the observed offset of the period-
metallicity envelopes for the period distributions of the field 
variables relative to those in clusters (Figs. 1 \& 4), and the 
presence or absence of the Oosterhoff gap in the clusters and in 
the field, we conclude that the origins of the two groups may be 
different. The presence in one and absence in the other of the 
Oosterhoff gap proves that the variation of the morphology of the 
HB with [Fe/H] differs between the two groups. This is another 
manifestation of the second parameter effect seen in only a 
handful of the Galactic globular clusters such as M13, NGC 7006, 
M2, NGC 5986 and the extreme cases of NGC 6388 and NGC 6441. 

     We agree with the growing consensus that the origin of at 
least a subset of the halo variables is the disruption by tidal 
friction of structures related to the dwarf spheroidals, where it 
is known that they have varying degrees of chemical evolution-
completion depending on their mass (luminosity). It is also known 
that the Oosterhoff gap characteristics varies among them. This 
variation is suggested to manifest itself in the difference in 
the morphology of the HB between the Galaxy and the dwarf 
spheroidals, similar to the differences seen between the bulk of 
the Galactic globular clusters and the second parameter clusters 
such as NGC 6838, NGC 6441, NGC 7006, M2, and NGC 5986. 

     8. A test of the hypothesis is to study the systematics of 
the morphology of the dwarf spheroidal HBs and its variation with 
metallicity. The HB morphology for them is expected to be 
systematically different from that in the Galaxy, and that 
difference is postulated to be a function of the mean metallicity. 
Such a study of comparative HB morphology between the normal and 
the second parameter globular clusters in the Galaxy and the 
local dwarf spheroidal companions is beyond the scope of this 
paper, but has been started.        

\acknowledgements ACKNOWLEDGEMENTS:  An early paper (unpublished), refereed by 
Geselle Clementini, has been rewritten here with a different goal 
and emphasis. The many suggestions by Clementini, leading to the 
old problems being analyzed from a new direction, are gratefully 
acknowledged. Other comments on the original paper by Fillipina 
Caputo were equally important in reconstituting it here, as were 
comments by Carla Cacciari and Flavio Fusi Pecci. I am grateful 
to them all. It is again a special pleasure to thank David 
Sandage for preparing the diagrams for press.

\clearpage

\section*{Figure Captions}

Figure 1.  Distribution of period with metallicity for the 302 type 
ab (fundamental mode) RR Lyrae variables studied by Layden (1994, 
1995). The full line [Eq. (1) of the text] is an approximation to 
the envelope of the shortest period variables at a given 
metallicity. The dashed line is Eq. (2) of the text for the 
log P values averaged over the narrow intervals of [Fe/H] listed 
in Table 1.

Figure 2.  The data from Fig. 1 for the Layden field star sample 
but with three linear envelope lines of Eqs. (3), (4) and 
(4') superposed. The envelope line is the same as in Fig. (1) for 
[Fe/H] more metal rich than -1.8 but two steeper envelopes from 
Eqs. (4) and (4') are shown as dashed lines for more metal poor 
stars.

Figure 3.  The same data for field stars as in Fig. 1 but with the 
parabolic fit of an envelope line from Eq. (5) superposed.

Figure 4.  The period-metallicity relation for 919 RR Lyrae stars in 
globular clusters from data in the Clement (2001) update of the 
Sawyer Hogg (1973) cluster variable catalog. Each cluster is 
represented by the horizontal line at its adopted metallicity 
from Harris (1996) on the metallicity scale of Zinn and West. The 
parabolic short-period locus for the field stars from Eq. (5) is 
shown. It fails to represent the short period distribution, 
requiring a shift of $\Delta$log P = 0.029 $\pm$ .007 toward longer 
periods for the cluster variables relative to those in the field.

Figure 5.  The predicted run of log T$_e$ with [Fe/H] keeping the 
luminosity fixed at the Catelan et al. (2004) calibration of Eq. 
(10), and using the four loci for the FBE  period-metallicity 
envelopes from Eqs. (1), (3), (4), (4') and (5), together with the 
Bono et al. mass equation from Eq. (15). These have been put into 
the pulsation Eq. (16) to generate the curves. Line number 1 
is for the linear envelope to the log P([Fe/H]) relation at the 
shortest periods in Fig. 1 (eq. 1); numbers 2 and 4 use the 
dashed lines in Fig. 2 from Eqs. (4) and (4') for [Fe/H] less 
than -1.8; number 3 is for the parabolic envelope of Eq. (5) from 
Fig. 3; number 5 is for the midpoint log P([Fe/H]) relation of 
Eq. (2) which is the dashed line in Fig. 1.

Figure 6.  Five predicted runs of log T$_e$ with [Fe/H] at the  
fundamental blue edge of the instability strip using a fixed 
parabolic envelope locus to the log P-[Fe/H] relation from Eq. 
(5) and Fig. 3, and five different RRL luminosity calibrations. 
Line 1 is for the linear luminosity calibration of Eq. (6) which 
was the S93a pulsation solution; line 2 is for the linear 
calibration of Clementini et al. (2003) from Eq. (12); line 3 is 
for the non-linear calibration of Eq. (7) by Caputo et al. 
(2000); line 4 is the non-linear calibration of Eq. (10) by 
Catelan et al. (2004); line 5 is for the two-line calibration of 
McNamana et al. (2004) of Eqs. (13) and (14).

Figure 7.  Predicted variation of B-V colors for the temperature 
curves in Fig. 5, using the Catelan et al. (2004) luminosity 
calibration in Eq. (10) and various assumed period-metallicity  
envelopes from Figs. 1-3. The conversion from the predicted 
temperatures in Fig. 5 to B-V is made using the adopted 
temperature-color relation of Eq. (18). Line 1 uses the linear 
envelope of equation 1, Fig. 1; lines 2 and 4 for [Fe/H] $<$ -1.8 
are for the dashed lines in Fig 2 which are Eqs. (4) and (4'); 
line 3 is for the parabolic envelope of Fig. 3, Eq. (5), and line 
5 is for the midpoint relation of Eq. (2).

Figure 8.  Predicted variation of B-V colors for the temperature 
curves in Fig. 6 using the fixed period-metallicity envelope of Eq. 
(5), (Fig. 3), and the five different luminosity calibrations of 
Eqs. (6), (8), (10), (12), (13), and (14). The adopted 
temperature-color relation is Eq. (18). The identification of the 
lines is the same as in the caption to Fig. 6.

Figure 9.  The points are the de-reddened photoelectric mean        
(B-V)$^0_{\rm mag}$ colors for 160 field RR Lyrae stars, plotted vs. 
Layden's measured [Fe/H] values, on the scale of Zinn and West. 
The line is the predicted run of color with metallicity from Eq. 
(19) calculated using the Catelan et al. (2004) luminosity 
calibration in Eq. (10) and the parabolic blue envelope locus to 
the period-metallicity relation of Fig 3, Eq. (5). The line is 
copied from Figs. 7 and 8.

Figure 10.  The de-reddened colors of RRab stars in globular 
clusters for which the number of RRab variables is 10 or greater. 
The Oosterhoff period gap is more prominent than in Fig. 4 because 
of this restriction on the number of variables. The metallicity 
scale is that of Zinn and West, as listed by Harris (1996). The 
E(B-V) reddening values of individual clusters are also from 
Harris. The colors, not listed here, are on the system of        
(B-V)$_{\rm mag}$, transformed to that system by Table 4 of Bono et al. 
(1995) if a different mean color definition was used in the 
original literature. The line is the prediction of Eq. (19), 
calculated using the luminosity calibration of Catelan et al. 
[eq. (10)], and the parabolic period-metallicity envelope of Eq. 
(5), together with the color transformation of Eq. (18).

Figure 11.  Combined data from Fig. 9 from field variables and from 
Fig. 10 for cluster variables. Crosses are the field stars. Dots 
are the cluster stars. The curve is from Eq. (20). It is 
similar to Eq. 19 for (Fe/H]) $>$ -2.0 but bends more strongly 
toward the red for more metal poor stars.

\clearpage
\begin{deluxetable}{cccccc}
\tablecaption{THE MID-POINT RIDGE LINE OF THE PERIOD-METALLICITY CORRELATION IN 
                       FIGURE 1}
\tablewidth{0pt}
\tablehead{
\colhead{Item} & \colhead{$\langle$[Fe/H]$\rangle$} & \colhead{Range of [Fe/H]} &
\colhead{n} & \colhead{$\langle$log P$\rangle$} & \colhead{(B-V)$^0_{\rm mag}$}\\
\colhead{(1)} &  \colhead{(2)}  & \colhead{(3)} & \colhead{(4)} & 
\colhead{(5)} & \colhead{(6)}\\
}
\startdata
mean &   -0.359 &  +0.07 to -0.70 &   21 &  -0.378  &  0.340\\
rms  &    0.049 &                 &      &   0.015  &  0.011\\
\\
mean &   -0.983 &  -0.71 to -1.19 &   16 &  -0.329  &  0.335\\
rms  &    0.039 &                 &      &   0.011  &  0.013\\
\\
mean &   -1.468 &  -1.20 to -1.79 &   71 &  -0.268 &   0.327\\
rms  &    0.019 &                 &      &   0.007 &   0.005\\
\\
mean &   -1.991 &   -1.80 to -2.49 &  34 &  -0.217 &   0.329\\
rms  &    0.031 &                  &     &   0.012 &   0.007\\
\\
mean &   -2.278 &   -2.06 to -2.49 &   7 &   -0.200 &  0.331\\
rms  &    0.059 &                  &     &    0.018 &  0.025
\enddata
\end{deluxetable}

\clearpage

\begin{deluxetable}{ccccl}
\tablecaption{CLUSTERS WHOSE RR LYRAE COLOR DATA ARE USED IN FIGS. 10 \&  11}
\tablewidth{0pt}
\tablehead{
\colhead{Cluster} & \colhead{[Fe/H]} & \colhead{E(B-V)} & \colhead{n$_{\rm ab}$}  &
\colhead{Literature} 
}
\startdata
 NGC 6362 &  -0.95 &  0.09 &  18 &  Olech et al. (2001)\\ 
 NGC 6712 &  -1.01 &  0.45 &   7 &  Sandage et al.(1966)\\  
 NGC 6171 &  -1.04 &  0.33 &  15 &  Dickens (1971), Sandage (S90a)\\
 NGC 6723 &  -1.12 &  0.05 &  23 &  Menzies (1974), (S90a)\\
 NGC 6121 &  -1.20 &  var  &  31 &  Sturch (1977), Cacciari (1979)\\
 NGC 1851 &  -1.22 &  0.02 &  21 &  Walker (1998)\\
 NGC 5904 &  -1.27 &  0.03 &  91 &  Broccato et al. (1996), Storm et al. (1991),\\
          &         &        &      &   Caputo et al. (1999)\\ 
 NGC 6981 &  -1.40 &  0.05 &  24 &  Dickens \& Flinn (1972), (S90a)\\
 NGC 6229 &  -1.43 &  0.01 &  30 &  Borissova et al. (2001)\\
 NGC 6934 &  -1.54 &  0.10 &  68 &  Kaluzny et al. (2001)\\
 NGC 5272 &  -1.57 &  0.01 & 145 &  Corwin \& Carney (2001)\\
 NGC 3201 &  -1.58 &  0.23 &  72 &  Piersimoni et al. (2002)\\ 
 IC 4499  &  -1.60 &  0.23 &  63 &  Walker \& Nemec (1996)\\
 NGC 7089 &  -1.62 &  0.06 &  17 &  Lee \& Carney (1999a)\\
 NGC 7006 &  -1.63 &  0.05 &  53 &  Wehlau et al. (1999)\\
 NGC 6809 &  -1.81 &  0.08 &   4 &  Olech et al. (1999)\\
 NGC 4590 &  -2.06 &  0.07 &  13 &  Walker (1994)\\
 NGC 5466 &  -2.22 &  0.00 &  13 &  Corwin et al. (1999)\\
 NGC 7078 &  -2.26 &  0.10 &  39 &  Bingham et al. (1984)\\
 NGC 5053 &  -2.29 &  0.04 &   5 &  Nemec (2004)\\
 NGC 6341 &   2.28 &  0.02 &  11 &  Carney et al. (1992)\\
\enddata
\end{deluxetable}

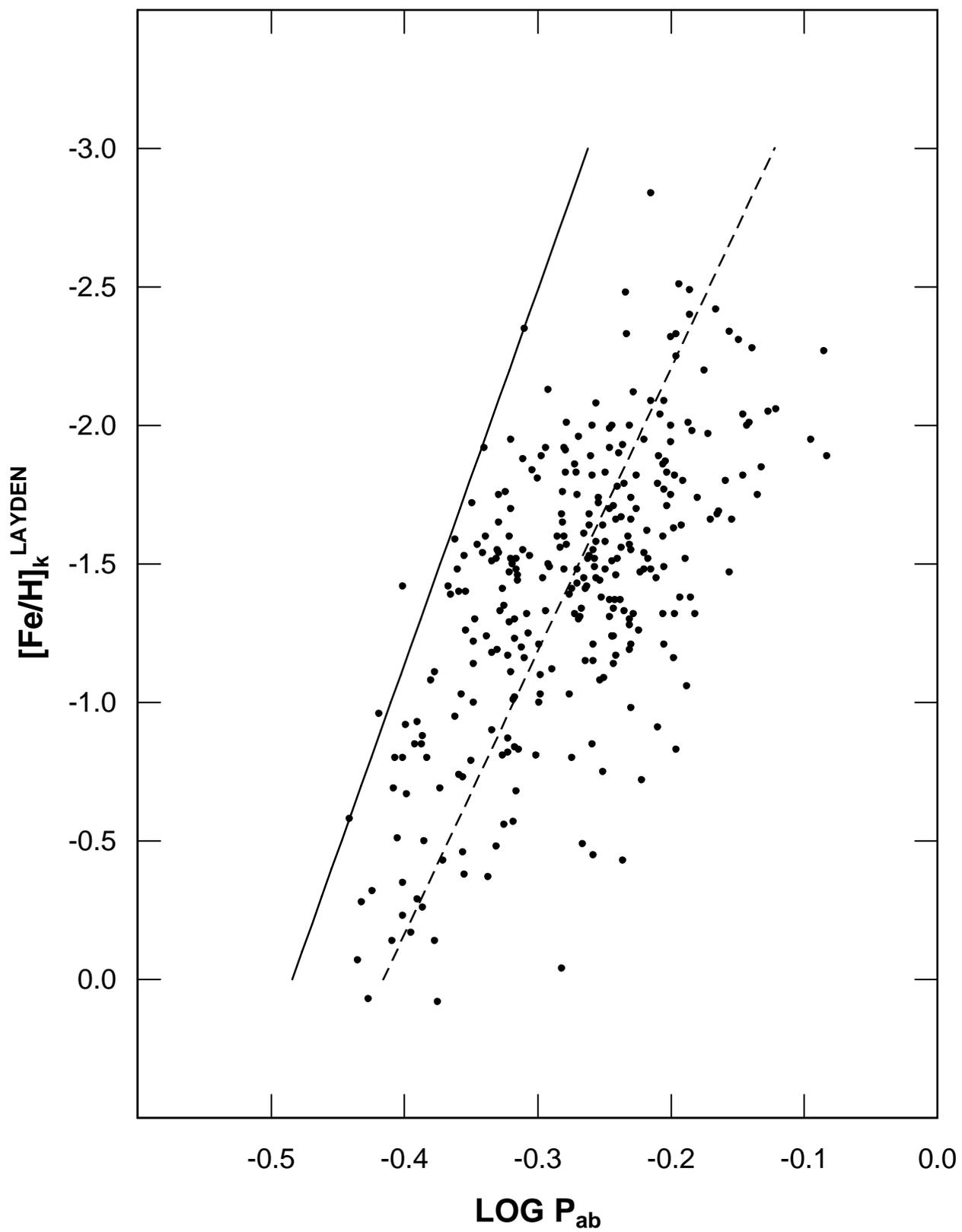
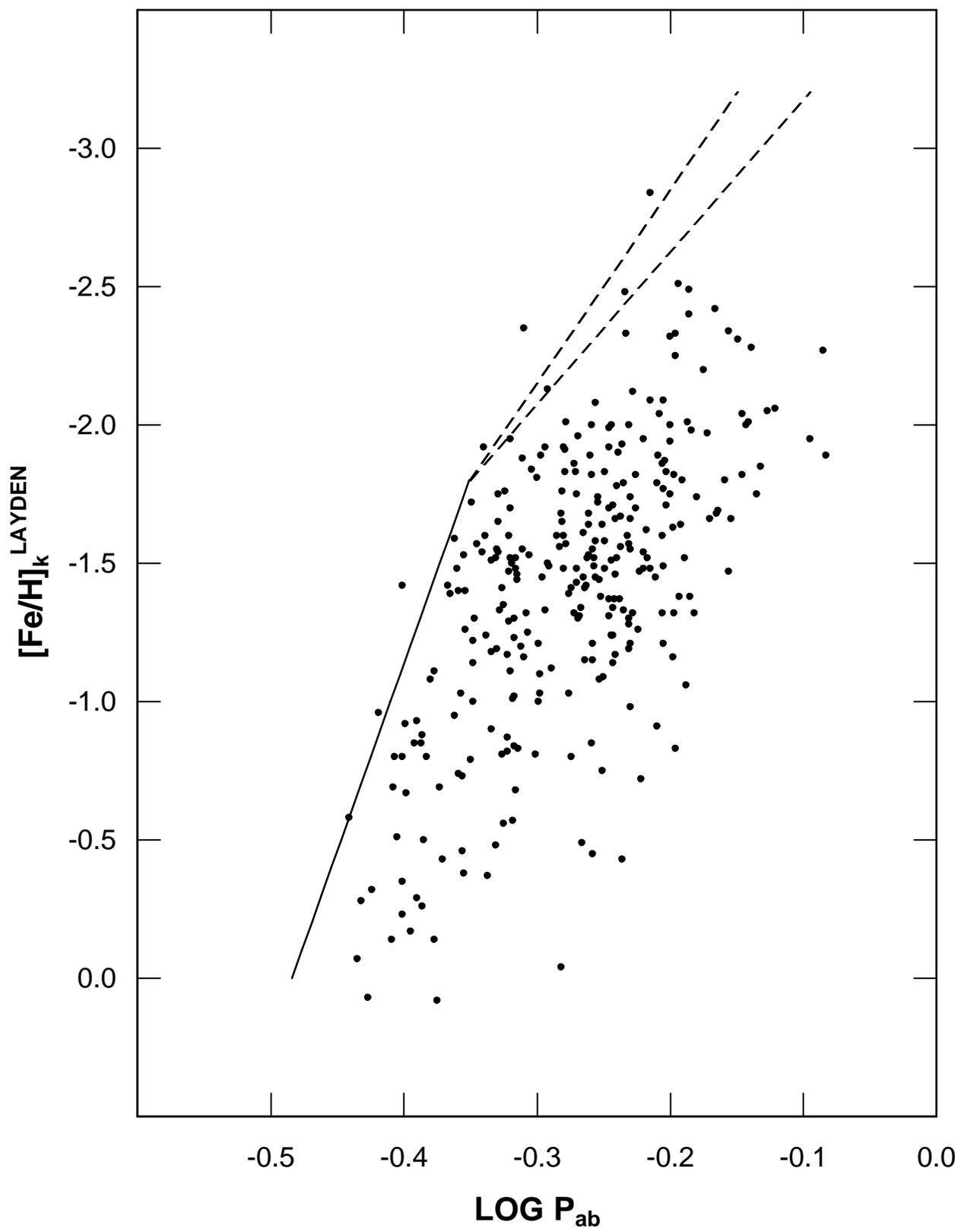
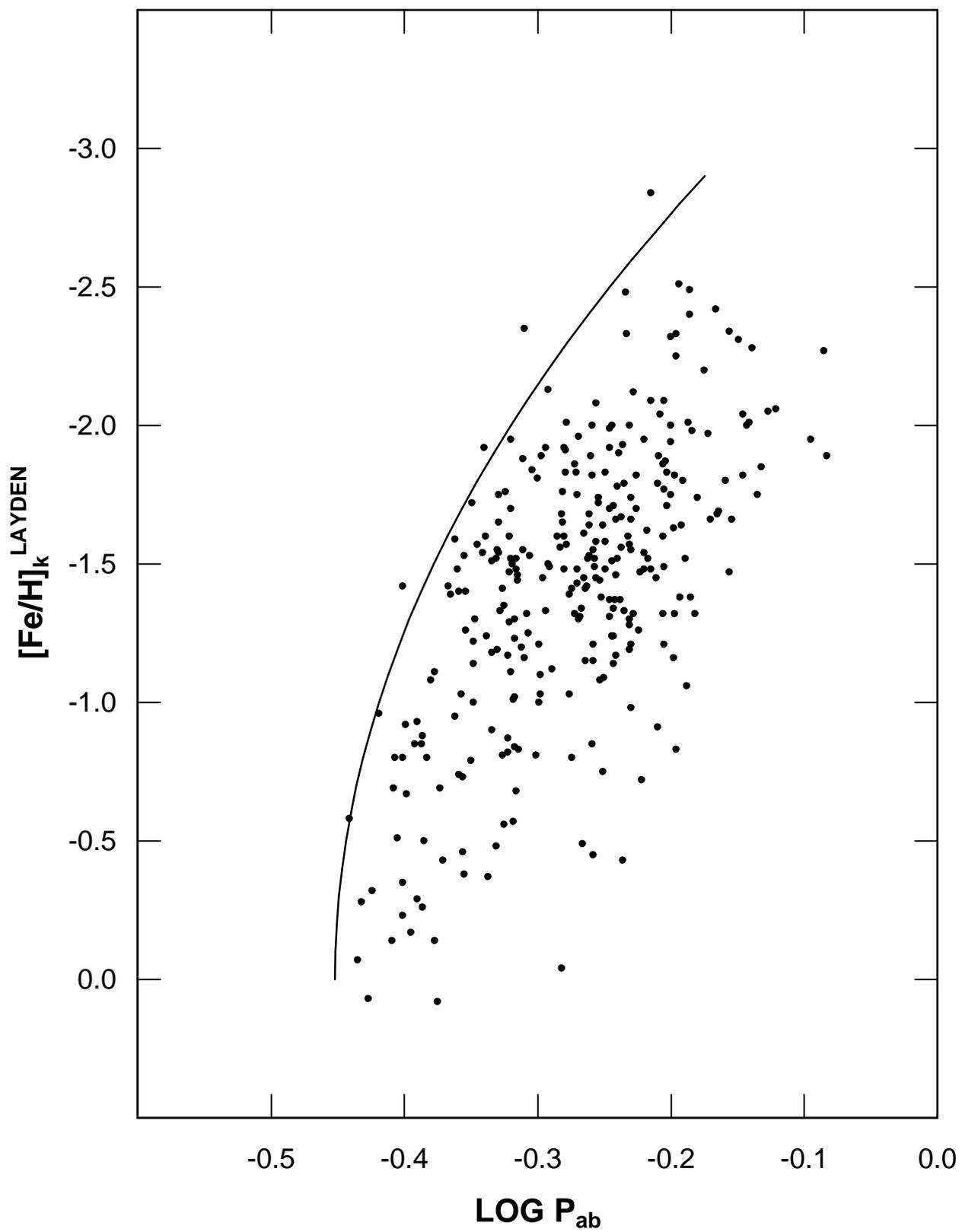
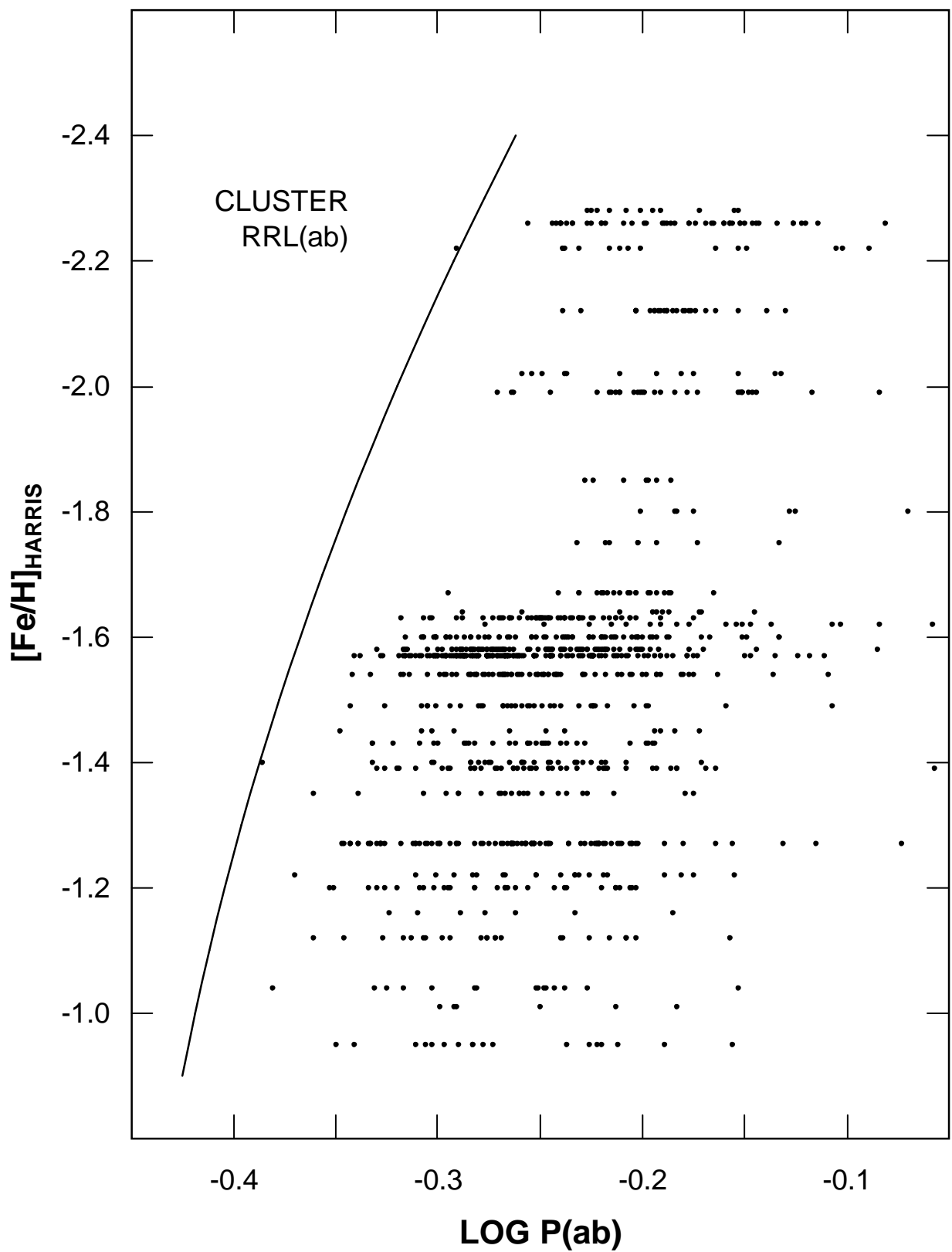
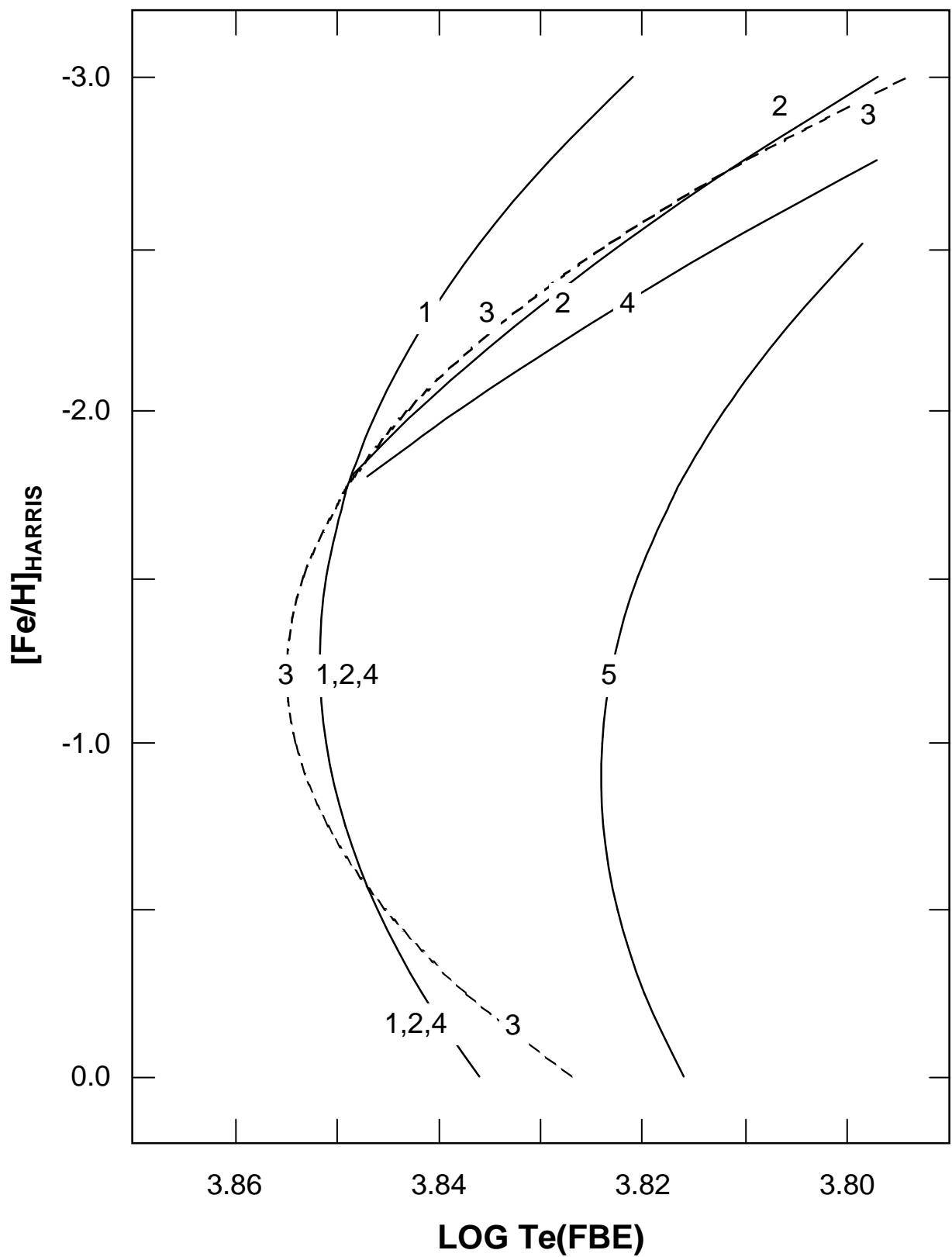
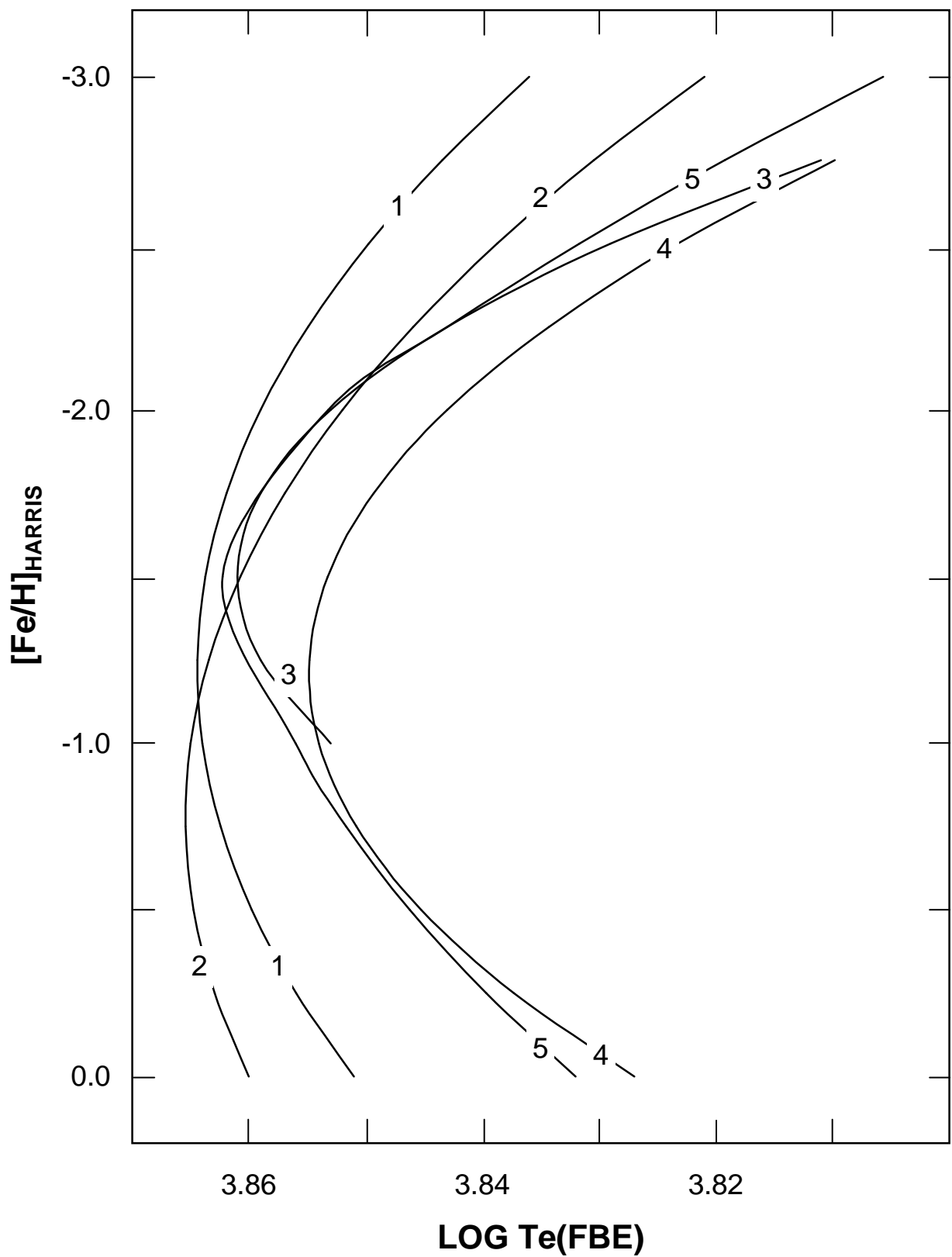
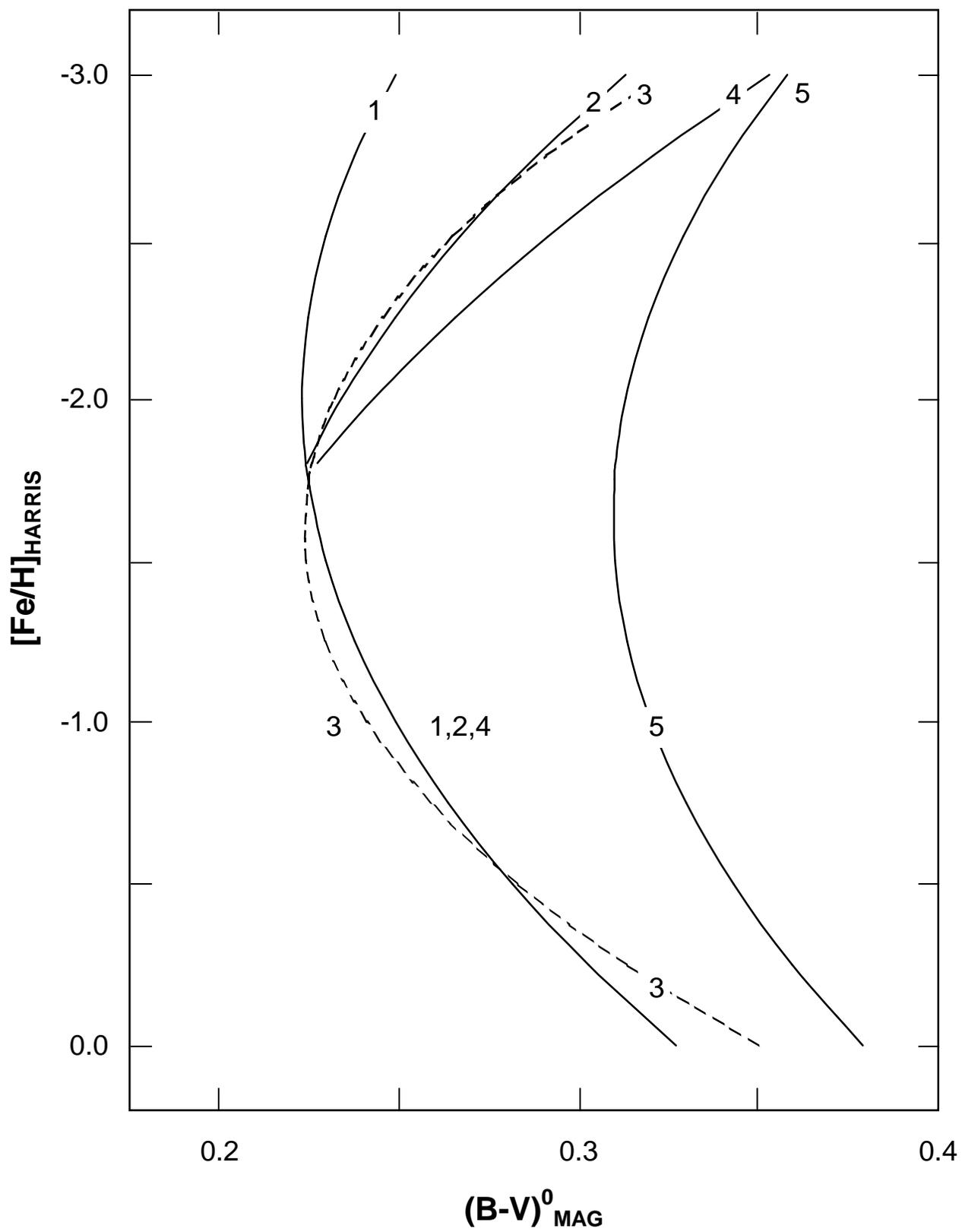
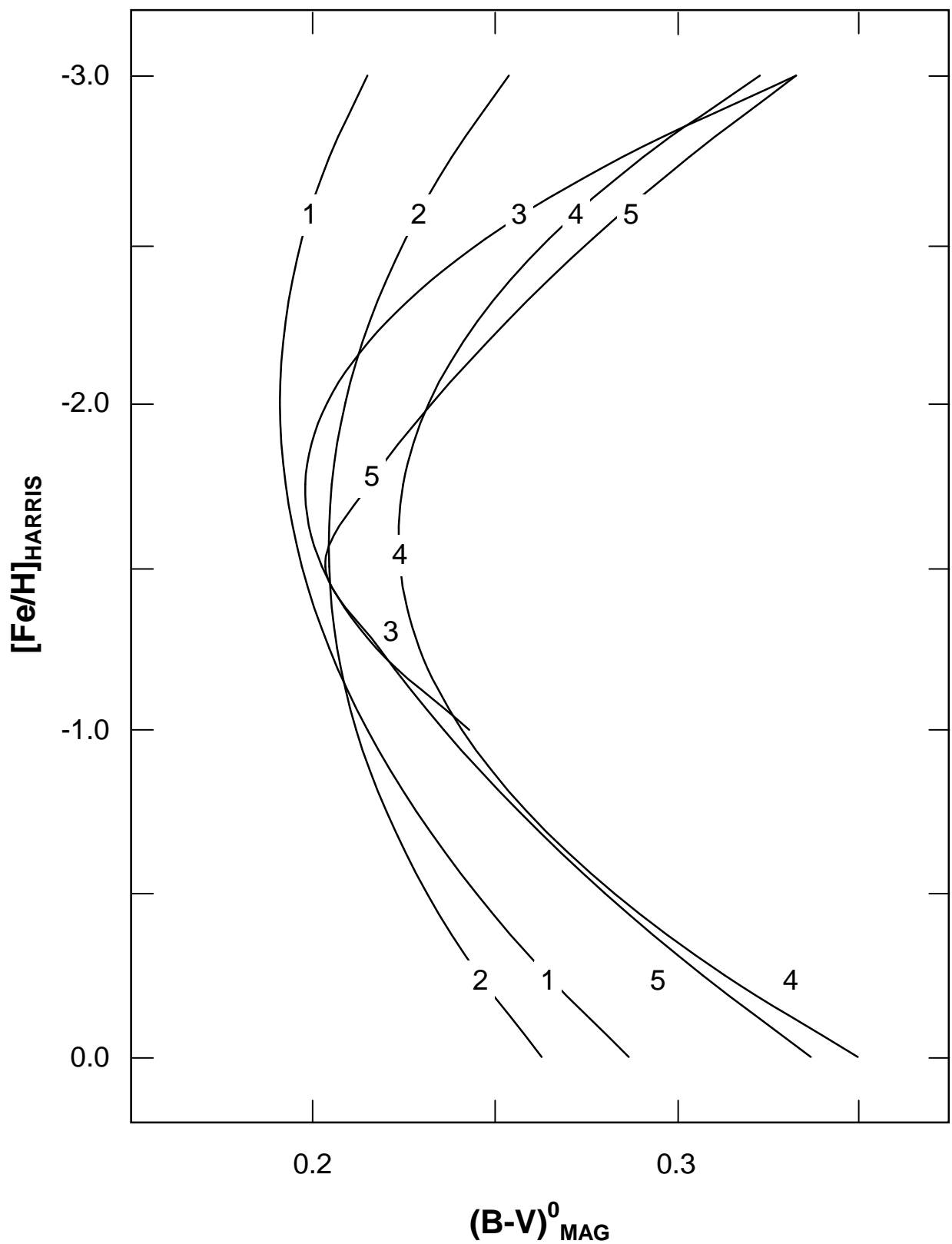
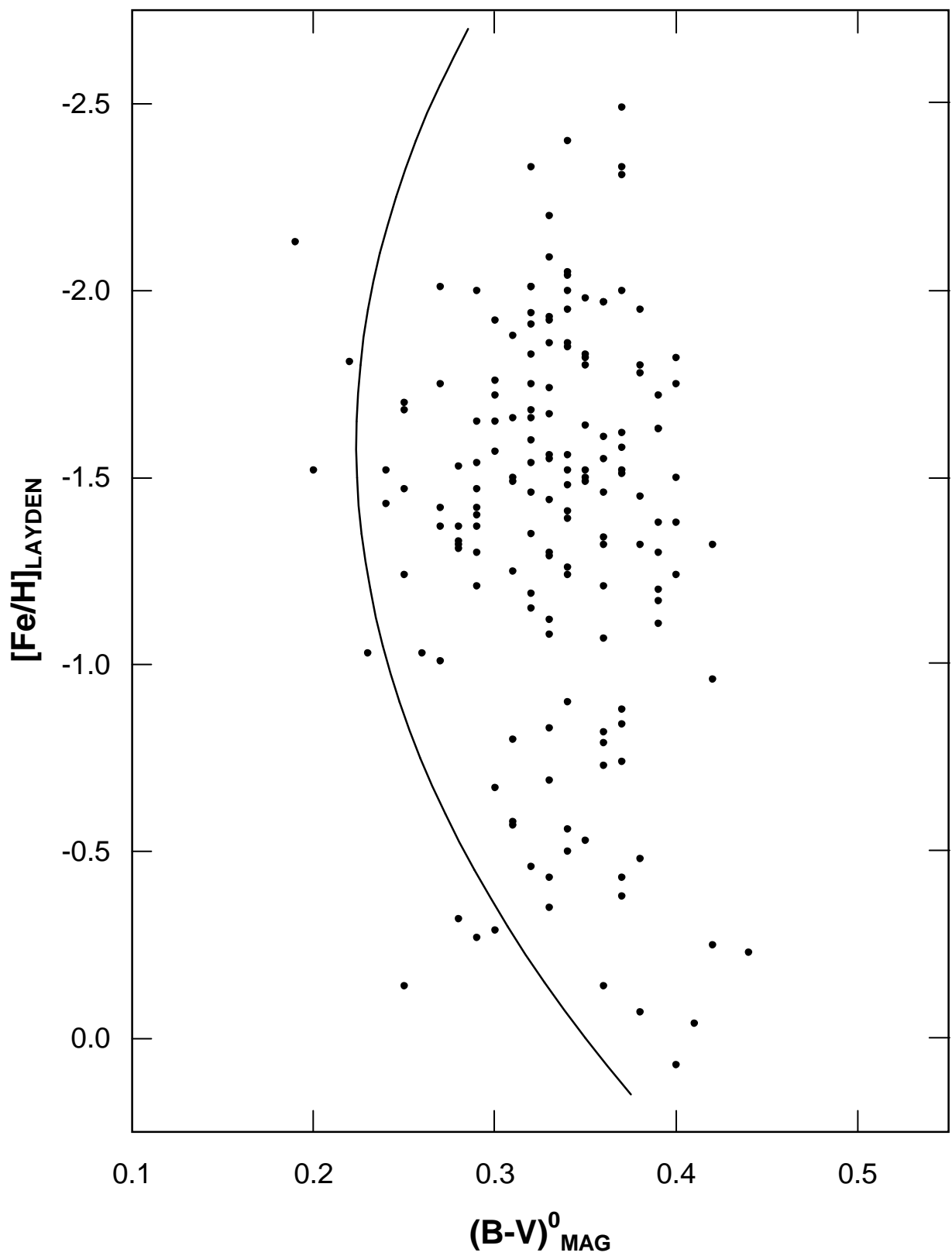
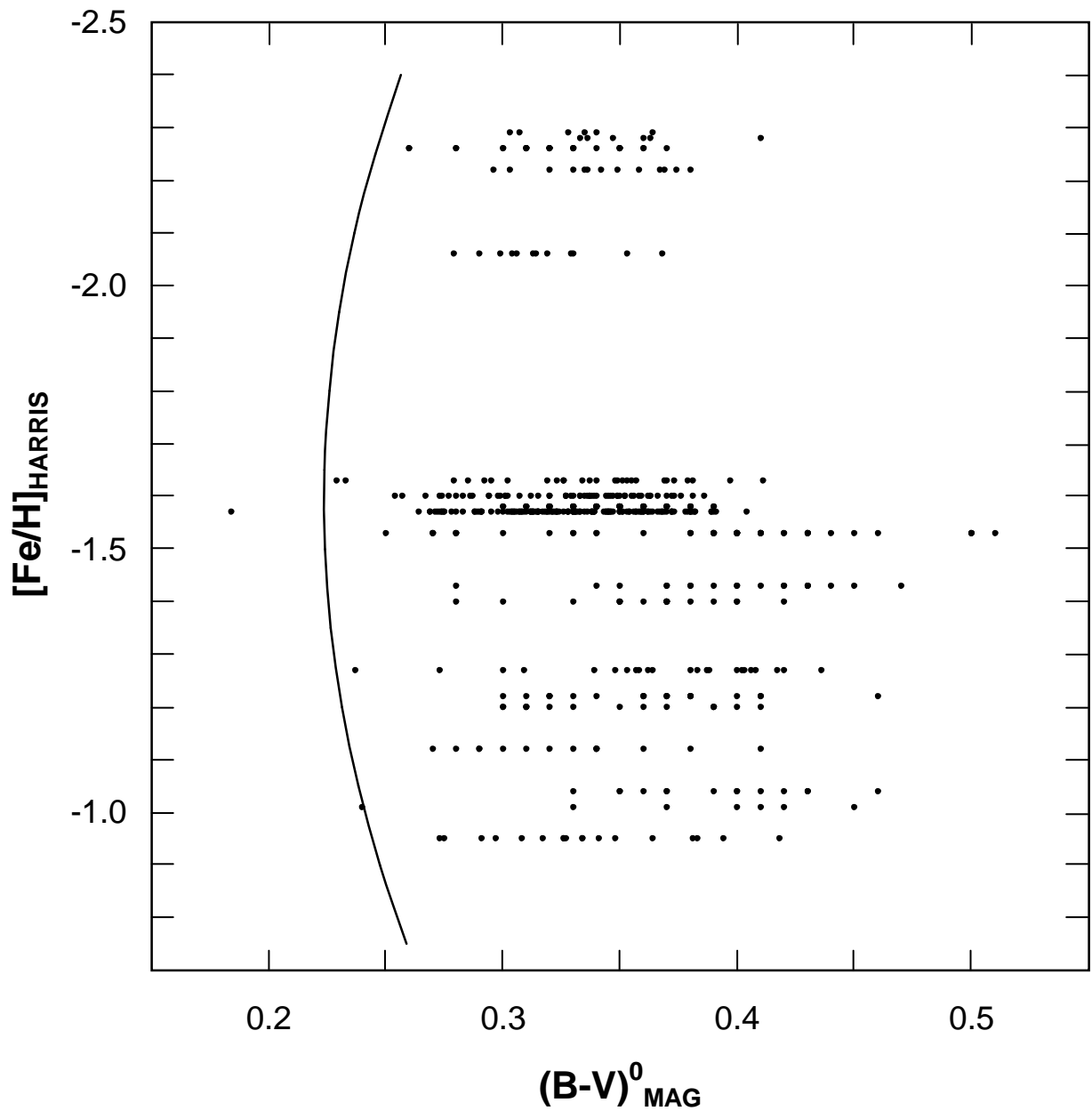
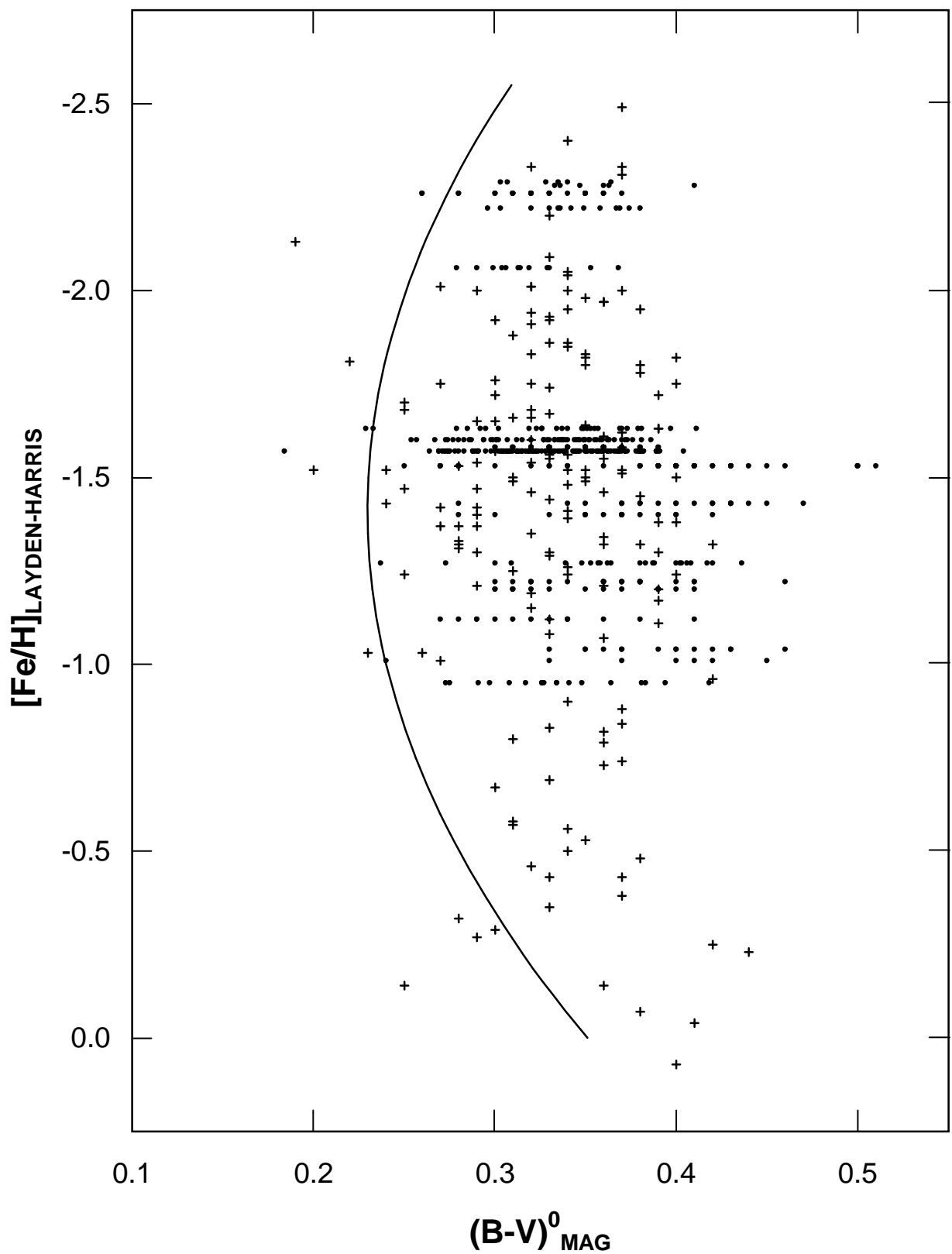
\end{document}